\documentclass[prb,twocolumn,superscriptaddress,showpacs]{revtex4-1}
\pdfoutput=1
\usepackage{graphicx}
\usepackage{amsfonts}
\def\kagome{kagome}

\def\abs#1{\vert #1 \vert}
\def\sw{f}
\def\Neg{<}
\def\Pos{>}

\begin{document}

\title{Order by disorder and phase transitions in a highly
frustrated spin model on the triangular lattice}

\author{A.\ Honecker}
\affiliation{Institut f\"ur Theoretische Physik,
  Georg-August-Universit\"at G\"ottingen, 37077 G\"ottingen, Germany}

\author{D.~C.\ Cabra}
\affiliation{Departamento de F\'{\i}sica, Universidad Nacional de La
Plata \\ and
Instituto de F\'{\i}sica La Plata, C.C.\ 67, (1900) La Plata, Argentina}

\author{H.-U.\ Everts}
\affiliation{Institut f\"ur Theoretische Physik, Leibniz Universit\"at Hannover,
  30167 Hannover, Germany}

\author{P.\ Pujol}
\affiliation{Laboratoire de Physique Th\'eorique,
  Universit\'e de Toulouse, UPS, (IRSAMC), 31062 Toulouse, France}
\affiliation{CNRS, LPT (IRSAMC), 31062 Toulouse, France}

\author{F.\ Stauffer}
\affiliation{RBS Service Recherche \& D\'eveloppement,
  Strasbourg -- Entzheim, 67836 Tanneries Cedex, France}

\date{August 26, 2011; revised November 14, 2011}

\begin{abstract}

Frustration has proved to give rise to an extremely rich phenomenology in 
both quantum and classical systems.  The leading behavior of the system 
can often be described by an effective model, where only the lowest-energy 
degrees of freedom are considered.  In this paper we study a system 
corresponding to the strong trimerization limit of the spin $1/2$ \kagome\ 
antiferromagnet in a magnetic field. It has been suggested that this 
system can be realized experimentally by a gas of spinless fermions in an 
optical \kagome\ lattice at $2/3$ filling. We investigate the low-energy 
behavior of both the spin $1/2$ quantum version and the classical limit of 
this system by applying various techniques. We study in parallel both 
signs of the coupling constant $J$ since the two cases display qualitative 
differences. One of the main peculiarities of the $J>0$ case is that, at 
the classical level, there is an exponentially large manifold of 
lowest-energy configurations. This renders the thermodynamics of the 
system quite exotic and interesting in this case. For both cases, $J>0$ 
and $J<0$, a finite-temperature phase transition with a breaking of the 
discrete dihedral symmetry group $D_6$ of the model is present. For $J<0$, 
we find a transition temperature $T^<_c/\abs{J} = 1.566 \pm 0.005$, {\it 
i.e.}, of order unity, as expected. We then analyze the nature of the 
transition in this case. While we find no evidence for a discontinuous 
transition, the interpretation as a continuous phase transition yields 
very unusual critical exponents violating the hyperscaling relation. By 
contrast, in the case $J>0$ the transition occurs at an extremely low 
temperature, $T^>_c \approx 0.0125 \, J$.  Presumably this low transition 
temperature is connected with the fact that the low-temperature ordered 
state of the system is established by an order-by-disorder mechanism in 
this case.

\end{abstract}


\maketitle


\section{Introduction}

The study of frustrated quantum magnets is a fascinating subject that has
stimulated many studies within the condensed matter
community in recent years.\cite{LMM11,SinzMisg,RSH04}
Such systems are assumed to be the main
candidates for a rich variety of unconventional phases and phase
transitions such as spin liquids and critical points with de-confined
fractional excitations.\cite{Science} Frustration can also play an
important r\^ole in classical systems. The phenomenon of order-by-disorder
\cite{OBD,CHS92} is the perfect example where the interplay of
frustration and fluctuations produces the emergence of unexpected
order. Order-by-disorder implies that 
a certain low-temperature configuration is favored by its high entropy, not by its 
low energy. Order-by-disorder can also occur in a 
quantum system, where a na\"ive argument suggests that quantum fluctuations
play the same r\^ole as thermal fluctuations in the classical system,
albeit there are  counterexamples where their r\^ole is in
fact quite different.\cite{kagomeus}

A particularly illustrative example
is provided by the spin $1/2$ antiferromagnet on the \kagome\ lattice.  A
spin gap appears to be present both at zero magnetization
\cite{SinzMisg,ZE90,LE93,lech97,waldt98,JWS08,YHW11,LSS11}
(see, however, Refs.~\onlinecite{SL09,NaSa11,Li11})
and at $1/3$ of the saturation value where it gives rise to a plateau
in the magnetization curve.
\cite{kagomeus,Hida01,CGHP02,SHSRS02,AHR04}
One would be tempted to believe that the nature of
the ground state is similar in both cases.
However, whether the ground state at zero field is ordered or not
is still under debate and also the existence of a plateau
in the isotropic spin 1/2 Heisenberg model
at magnetization $1/3$ has been questioned recently.\cite{NaSa10,SaNa11}
Nevertheless, the existence of a plateau at magnetization $1/3$ is
quite clear for easy-axis exchange anisotropies\cite{kagomeus,CGHP02}
and, using a correspondence with a quantum dimer model on the honeycomb
lattice,\cite{MS01}
the ground state is identified as an ordered
array of resonating spins.\cite{kagomeus,kagomeus1}

In this paper we study an effective model that arises in the strong
trimerization limit of the spin $1/2$ \kagome\ antiferromagnet.\cite{sub95}
This model has played an important r\^ole in analyzing the zero-field
properties of the \kagome\ antiferromagnet,\cite{mila98,mami00} but
here we will focus on magnetization $1/3$ of the Heisenberg model,
corresponding to full polarization of the physical spin degrees in the effective
model. Thus, we are left with the chirality degrees of freedom of the
original antiferromagnet which we will treat as `spin' variables.
In this sense our spin system can be considered as a
purely orbital model similar to compass models recently considered in the
literature (see, e.g., Refs.\
\onlinecite{NuFr05,DBM05,TI07,WJ08,SWT09,WJ09,TOH10,WJL10,OH11}).
As an experimental realization of this
model a system of spin-polarized fermions trapped in a trimerized optical 
\kagome\ lattice at $2/3$ occupancy has been suggested. 
\cite{SBCEFL04,DEHFSL05,DFEBSL05} 
In fact, experimental realization of an optical \kagome\ lattice
has been reported recently,\cite{JGTHVS11} albeit using a setup
which does not allow direct control of trimerization.

Beyond the particular realizations of our model its very rich physics 
which results from the interplay between classical and quantum fluctuations 
and frustration makes it an interesting model in its own right. As will be 
shown in this paper, a Hamiltonian with an (unusual)  discrete symmetry    
but with a continuous degeneracy of the classical ground state, as it would 
be expected for a Hamiltonian with a continuous symmetry, is just one aspect 
of the rich phenomena emerging from this model.

The present paper is organized as follows: in section \ref{sec:Hsym} we present the
Hamiltonian and the symmetries of the classical and spin $1/2$ cases. The
Hamiltonian can be defined for both signs of the coupling constant $J$. We
deliberately discuss in parallel the two cases throughout the entire
paper to point out their similarities and differences. The spin $1/2$ case
is then treated in section \ref{secED} by means of exact diagonalization
techniques, and we argue that a finite-temperature phase transition takes
place. Since exact diagonalization can access only small lattices,
we move to the classical model in section \ref{secCl}. We study in detail
the manifold of lowest-energy configurations and their corresponding
spin-wave spectra. The effect of soft modes in the order-by-disorder
selection mechanism
is argued to be the origin for the phase transition of the $J>0$ case, in
contrast to the $J<0$ case, where the transition is of a more conventional
purely energetic origin.
In section \ref{secMC} we apply Monte-Carlo techniques to the classical
model and determine the transition temperature for $J>0$ and $J<0$.
We also analyze the universality class of the transition, however, only
for $J<0$ since the transition temperature for $J>0$ turns out to be
so low that it is difficult to access.
Finally section \ref{sec:Concl} is devoted to
some concluding remarks and comments.

\section{Hamiltonian and symmetries}

\label{sec:Hsym}

\subsection{Hamiltonian}

\begin{figure}[tb!]
\leftline{(a)} 
\vspace*{-3mm}
\includegraphics[width=0.75\columnwidth]{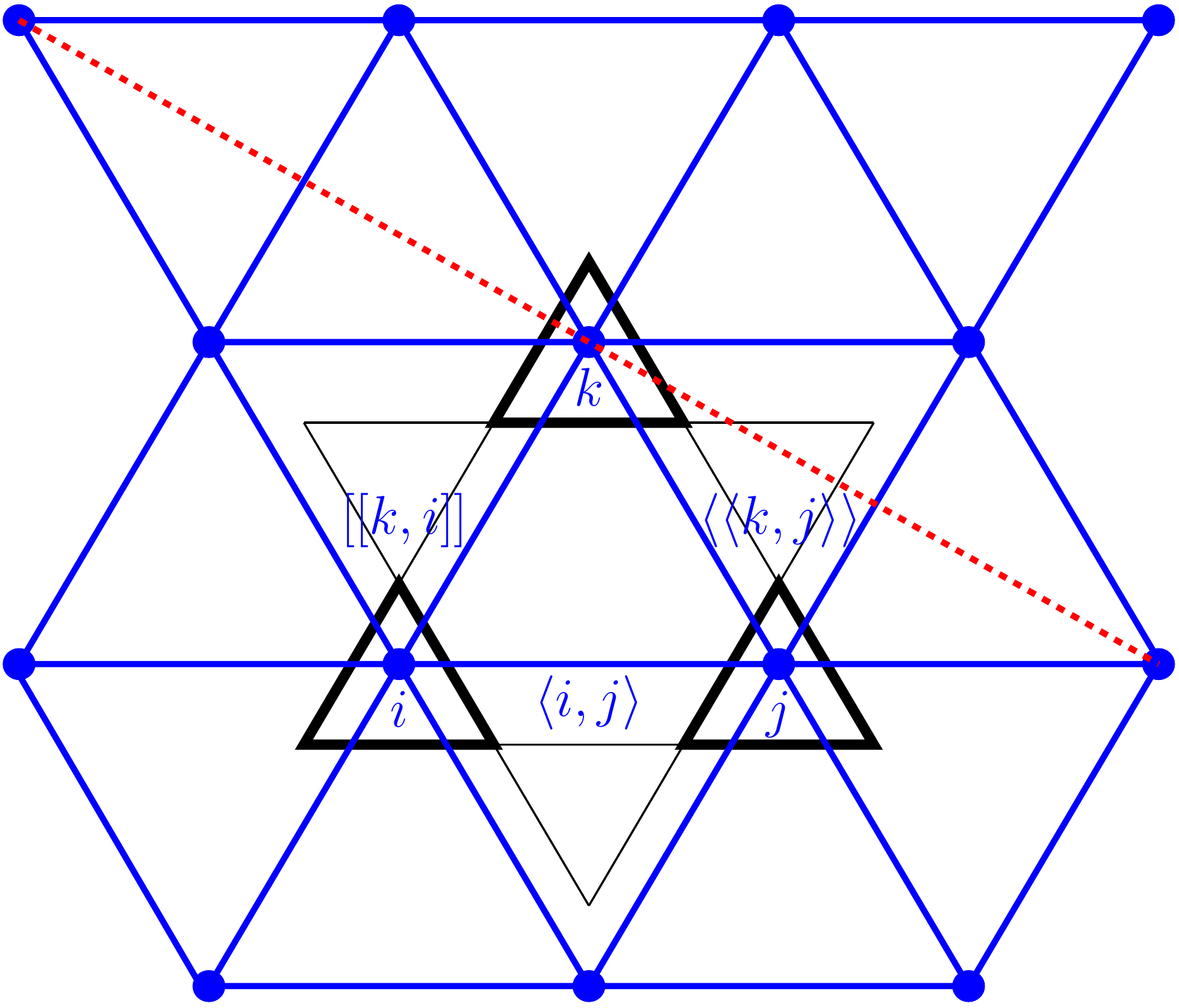}

\vspace*{5mm}
\leftline{(b)} 
\vspace*{-1mm}
\includegraphics[width=0.9\columnwidth]{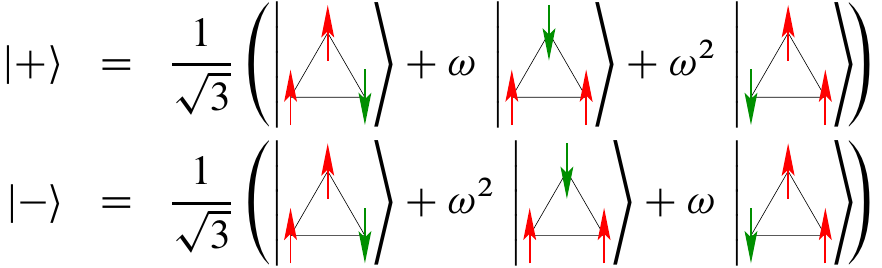}

\vspace*{5mm}
\leftline{(c)} 
\vspace*{-3mm}
\includegraphics[width=0.8\columnwidth]{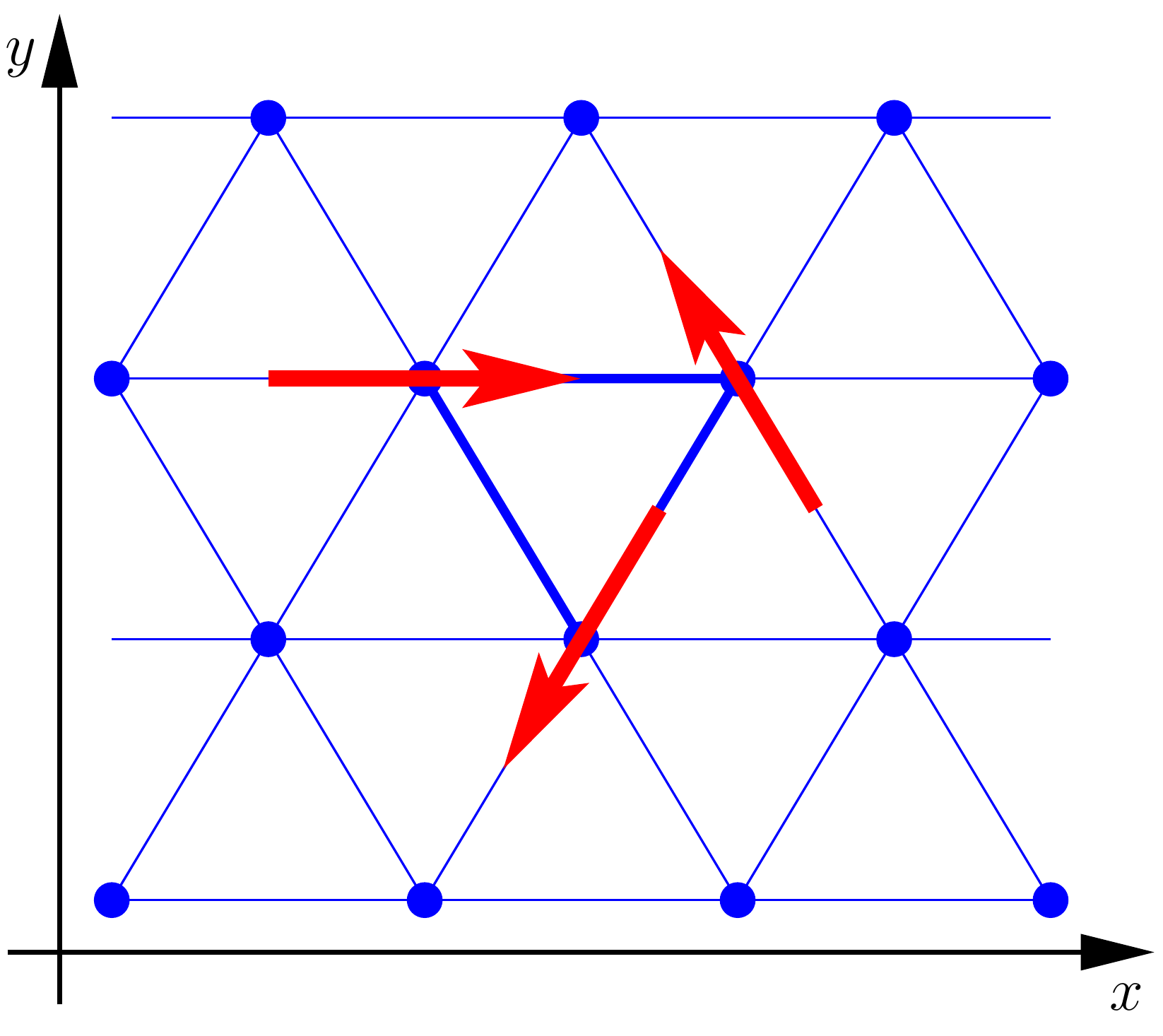}
\caption{(Color online)
(a) Triangular lattice with assignment of bonds
to the three different directions and underlying trimerized \kagome\
lattice.
(b) The two chirality states of a triangle.
(c) Assignment of the vectors $\vec{e}_i$ to the bonds of the
triangular lattice for the alternative representation
(\ref{eqHalt}) of the Hamiltonian.
}
\label{figLattice}
\end{figure}

We will study the Hamiltonian
\begin{equation}
H = J \, \left(\sum_{\langle i,j\rangle} T^A_i T^C_j
+ \sum_{\langle\langle k,j \rangle\rangle} T^A_k T^B_j
+ \sum_{[[ k,i ]]} T^C_k T^B_i \right) \, ,
\label{eqH}
\end{equation}
where
\begin{eqnarray}
T^A_i &=& S^+_i + S^-_i = 2\, S^x_i \, , \nonumber \\
T^B_i &=& \omega \, S^+_i + \omega^2 \, S^-_i
= -S^x_i - \sqrt{3} \, S^y_i \,  \, , \label{eqTop} \\
T^C_i &=& \omega^2 \, S^+_i + \omega \, S^-_i
= -S^x_i + \sqrt{3} \, S^y_i \, , \nonumber
\end{eqnarray}
with the third root of unity $\omega = {\rm e}^{2 \pi i/3}$.
The sums in (\ref{eqH}) run over the bonds of a triangular lattice,
each corresponding to one of the three distinct directions of the
lattice, as sketched in Fig.~\ref{figLattice}(a).

The Hamiltonian (\ref{eqH}) arises as an effective Hamiltonian
for the trimerized \kagome\ lattice, sketched in Fig.~\ref{figLattice}(a)
behind the triangular lattice.
Our notation follows the derivation from the half-integer spin Heisenberg
model for the case where the remaining magnetic degrees of freedom are
polarized.\cite{sub95} In this case, there are two pseudo-spin states
of opposite chirality for each triangle, see Fig.~\ref{figLattice}(b).
As reviewed in appendix \ref{sec:AppDeriv},
plain first-order perturbation theory of the ${\cal S}^z$-${\cal S}^z$ interactions between
two triangles yields Eq.~(\ref{eqH}) where the exchange constant $J$
is proportional to the inter-triangle exchange constant of the \kagome\
lattice and would thus typically assumed to be antiferromagnetic
($J>0$). Note that we have chosen a convenient normalization of $J$.
A similar derivation starting  from a Fermi gas with two atoms
per trimer also leads to the Hamiltonian (\ref{eqH}).\cite{SBCEFL04}

Due to the two possible chiralities on each
triangle, the pseudo-spin operators $\vec{S}_i$ should be considered as
quantum spin-1/2 operators. The derivations\cite{sub95,SBCEFL04} also
suggest a positive $J>0$ to be more natural.
In this paper we will relax these constraints and, for reasons that
will become clear later, consider
also classical spins, {\it i.e.}, unit vectors  $\vec{S}_i$, and
the case $J < 0$.

Note that the Hamiltonian (\ref{eqH}) is
not symmetric under reflections of the lattice. Our conventions agree
with those of Ref.\ \onlinecite{sub95} where this Hamiltonian appeared first,
while some more recent works\cite{SBCEFL04,DEHFSL05,DFEBSL05,Z05} use a
reflected convention for the chirality.
Note furthermore that our conventions for $J$ differ by a factor
4 from previous studies of the model (\ref{eqH}).\cite{DEHFSL05,DFEBSL05}

It may also be useful to represent the Hamiltonian (\ref{eqH}) in a more
compact form\cite{hfm2006}
\begin{equation}
H = 4\,J \, \sum_{\langle i,j\rangle}
\left(\vec{e}_i\cdot \vec{S}_i\right) \, \left(\vec{e}_j\cdot \vec{S}_j\right)
\, ,
\label{eqHalt}
\end{equation}
where the vectors $\vec{e}_i$ are indicated in Fig.~\ref{figLattice}(c):
for each bond one has to choose $\vec{e}_i$ and $\vec{e}_j$ as in the
corresponding bond of the bold triangle. For example, for each horizontal
bond $\langle i,j\rangle$, one needs to choose $\vec{e}_i = \pmatrix{1 \cr 0}$ for the left
site $i$ and $\vec{e}_j = \frac{1}{2} \, \pmatrix{-1 \cr \sqrt{3}}$ for the
right site $j$.

Models which are very similar to (\ref{eqH}) have recently
been studied in the context of spin-orbital models (see, e.g., Refs.\
\onlinecite{NuFr05,DBM05,TI07,WJ08,SWT09,WJ09,TOH10,WJL10,OH11}).

\subsection{Symmetries}

The Hamiltonian (\ref{eqH}) has the following symmetries on an
infinite lattice:
\begin{enumerate}
\item Translations $T_x$, $T_y$ along the two fundamental directions
  of the lattice.
\item Simultaneous rotation $R_{2 \pi/3}$ of the lattice
  and all spins around the $z$-axis by angles of $2 \pi/3$
  (the latter rotation amounts to a cyclic exchange of $T^A_i$, $T^B_i$,
  and $T^C_i$).
\item A rotation by $\pi$ around the $z$-axis in spin space: 
  $P:\quad S^x_i \mapsto -S^x_i,\ S^y_i \mapsto -S^y_i$ while keeping the
  lattice fixed.
\item A spatial reflection combined with
  rotation of all spins around a suitable axis in the $x$-$y$-plane
  by an angle $\pi$. One particular choice is
  $I:\quad S^x_i \mapsto S^x_i,\ S^y_i \mapsto -S^y_i,\ S^z_i \mapsto -S^z_i$,
  combined with a reflection of the lattice along the dashed line in
  Fig.~\ref{figLattice}(a).
\item For spin 1/2, there is another symmetry implemented by
\begin{equation}
Q = \prod_i \left(2 \, S^z_i \right) \, .
\label{eqQdef}
\end{equation}
  Conservation of $Q$ means that the number of spins pointing up
  (or down) along the $z$-axis is a good quantum number modulo two.
  This conservation law is most easily verified by observing that the
  interaction terms in (\ref{eqH}) always invert a pair of spins in an
  eigenbasis of $S^z$.
\end{enumerate}
The choice of factors in (\ref{eqQdef}) ensures that $Q^2 = {\mathbf 1}$.
Furthermore, one has $R_{2 \pi/3}^3 = P^2 = I^2 = {\mathbf 1}$.
$R_{2 \pi/3}$ and $P$ together generate the abelian group
${\mathbb Z}_6 \cong {\mathbb Z}_3 \times {\mathbb Z}_2$, as
described for instance in Ref.~\onlinecite{DEHFSL05}. The combination of
$R_{2 \pi/3}$, $P$, and $I$ generates the dihedral group $D_6$, which
is non-abelian ($I \, R_{2 \pi/3} \, I = R_{2 \pi/3}^{-1}$).
Finally, $R_{2 \pi/3}$ and $I$ generate the symmetric group $S_3$
which can be traced to the point-group symmetry of the underlying
\kagome\ lattice. The operators $R_{2 \pi/3}$, $P$, and $I$ 
leave the Hamiltonian (\ref{eqH}) invariant irrespective of the value
of the spin quantum number. Thus, the group $D_6$ is a symmetry also
of the classical variant of the model.

The symmetries $P$ and $Q$ are not present in the underlying
\kagome\ lattice, hence they should be specific to the lowest-order
approximation.\cite{sub95,SBCEFL04} Indeed, at least in the derivation
from the Heisenberg model one observes that already the next
correction\cite{Z05} breaks the symmetries $P$ and $Q$.

Now let us consider the consequences of the combination of $I$ and $Q$
for the spin-1/2 model
on a finite lattice with $N$ sites. Then the relation
$I \, S^z_j \, I = -S^z_j$ leads to
\begin{equation}
I \, Q \, I = (-1)^N \, Q  \, .
\label{eqIQrel}
\end{equation}
Since the eigenvalues of $Q$ are $q = \pm 1$,
this implies that $I$ is an isomorphism
between the subspace with $q=-1$ and the subspace with $q=1$ for odd $N$
and spin 1/2.

\section{Quantum system: exact diagonalization for spin 1/2}

\label{secED}

In this section we will present numerical results for the Hamiltonian
(\ref{eqH}) with spin 1/2. We impose periodic boundary conditions
and use the translational symmetries $T_x$, $T_y$
in order to classify the states
by a momentum quantum number $\vec{k}$. We only consider lattices
which do not frustrate a potential three-sublattice order, {\it i.e.},
only values of $N$ that are multiples of three. For the system sizes $N$
considered already in Refs.~\onlinecite{DEHFSL05,DFEBSL05},
we will use the same lattices. In particular, the
$N=12$, $21$, and $24$ lattices are shown in Fig.~9 of
Ref.~\onlinecite{DFEBSL05}. Furthermore, we will
consider the $N=27$ lattice which can be found, e.g., in Fig.~1
of Ref.~\onlinecite{NiNa88}.

Let us briefly discuss the consequences of the other symmetries
mentioned above. We did not make explicit use of $P$, although
it is present for any lattice. However, the symmetry $Q$
(which is also present for any lattice, see (\ref{eqQdef}))
is easily implemented if we work in an $S^z$-eigenbasis.
Concerning the rotation $R_{2 \pi/3}$,
it is not possible to find lattices
for all $N$ such that it is a symmetry.
If  $R_{2 \pi/3}$ is a symmetry,
we use it to select one representative $\vec{k}$
for all equivalent momenta. Finally, the presence of the symmetry $I$ is
more delicate. We have performed computer checks and found that
most of the lattices under consideration have a suitable spatial
reflection symmetry, ensuring that $I$ is a symmetry. The only
exception is the $N=21$ lattice where there is no such
reflection symmetry.
Nevertheless, we find the same spectra in the subspaces with
$q=-1$ and $q=1$ also for $N=21$. Therefore, for $N$ odd we can
choose representatives for all symmetry sectors in the subspace with $q=1$.

For $N \le 21$ the translational symmetries and $Q$
lead to matrices with dimension up to
49~940 and we can obtain all eigenvalues. Dimensions increase up to
2~485~592 for $N=27$. In this case, we have used the Lanczos method
to compute the $n$ lowest eigenvalues in each sector. Mainly for reasons
of CPU time, we restrict to $n \approx 70$ ($150$) for $N=27$ ($24$)
and $J > 0$.

\subsection{Low-lying spectra}

\begin{figure}[tb!]
\begin{center}
\includegraphics[width=\columnwidth]{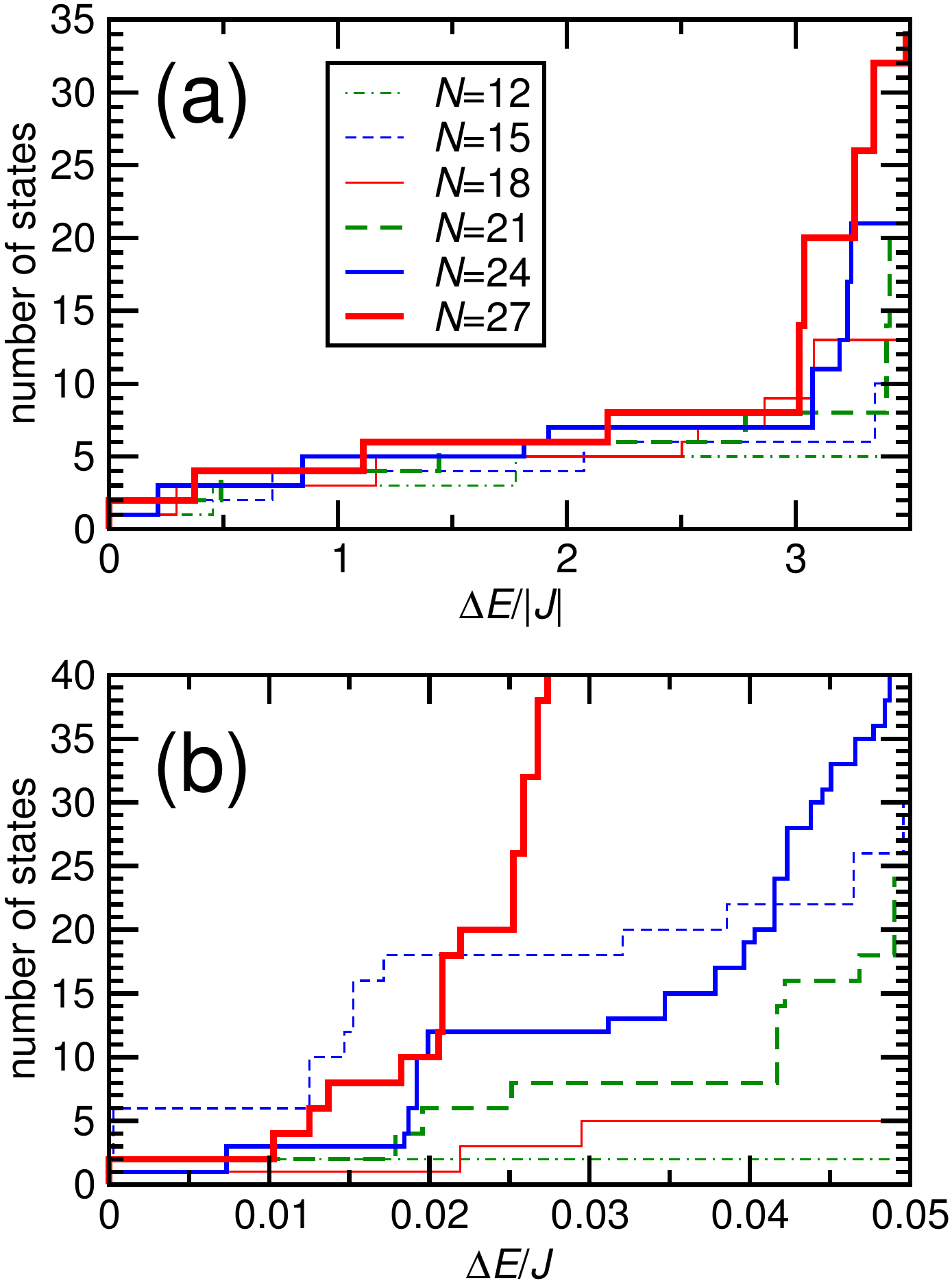}
\end{center}
\caption{(Color online) Integrated density of states, {\it i.e.}, number of states with energy less or equal than $\Delta E$ above the
ground state, for (a) $J < 0$ and (b) $J>0$.}
\label{figSpec}
\end{figure}

Let us first look at the spectra. In order to take degeneracies
into account, in Fig.~\ref{figSpec} we show the integrated density
of states, {\it i.e.}, the number of
states with energy less or equal than $\Delta E$ above the
ground state.

Panel (a) of Fig.~\ref{figSpec} shows results for $J < 0$.
These results extend previously presented results\cite{DFEBSL05}
for $N=12$, $18$, and $21$ to higher energies and larger $N$.
One observes that there are at most 8 states for energies
$\Delta E \lesssim 3 \, \abs{J}$ with a substantial density of states
setting in at higher energies. This suggests a thermodynamic
gap $\approx 3\, \abs{J}$.

Fig.~\ref{figSpec}(b) shows the density of states for $J > 0$,
extending previously published results for $N=18$, $21$,
and $24$.\cite{DEHFSL05,DFEBSL05} In this case, we observe
a large density of states at substantially lower energies than for
$J<0$. This large density of states is reminiscent of the large
density of non-magnetic excitations observed in the spin-1/2
Heisenberg antiferromagnet on the \kagome\ lattice, both at zero
magnetic field\cite{lech97,waldt98} and on the one-third
plateau.\cite{kagomeus}
In particular the $N=27$ data presented in Fig.~\ref{figSpec}(b)
shows a large density of states for $\Delta E \gtrsim 0.02\,J$.
On the other hand, one observes at most 8 states with
$\Delta E \lesssim 0.012 \, J$ in Fig.~\ref{figSpec}(b)
for a given system size $N$. From these observations we infer that
a gap is at most on the order of $0.02\,J$ if present at all.

Since an ordered ground state breaks the symmetry group $D_6$, such
a ground state should be six-fold degenerate. Indeed, classical
and semiclassical considerations predict a six-fold degeneracy
in an ordered state (see section \ref{secCl} below). However,
there is no clear separation of 6 low-lying states from the
remainder of the spectrum 
for $J<0$ (see Fig.~\ref{figSpec}(a)),
and even less so for $J>0$ (Fig.~\ref{figSpec}(b)). The considered
lattice sizes may be too small to observe the expected low-energy
structure of the spectrum. However, correlation functions exhibit
pronounced $120^\circ$ correlations already
on these small lattices.\cite{DEHFSL05,DFEBSL05}

\subsection{Specific heat}

\label{secCq}

\begin{figure}[tb!]
\begin{center}
\includegraphics[width=\columnwidth]{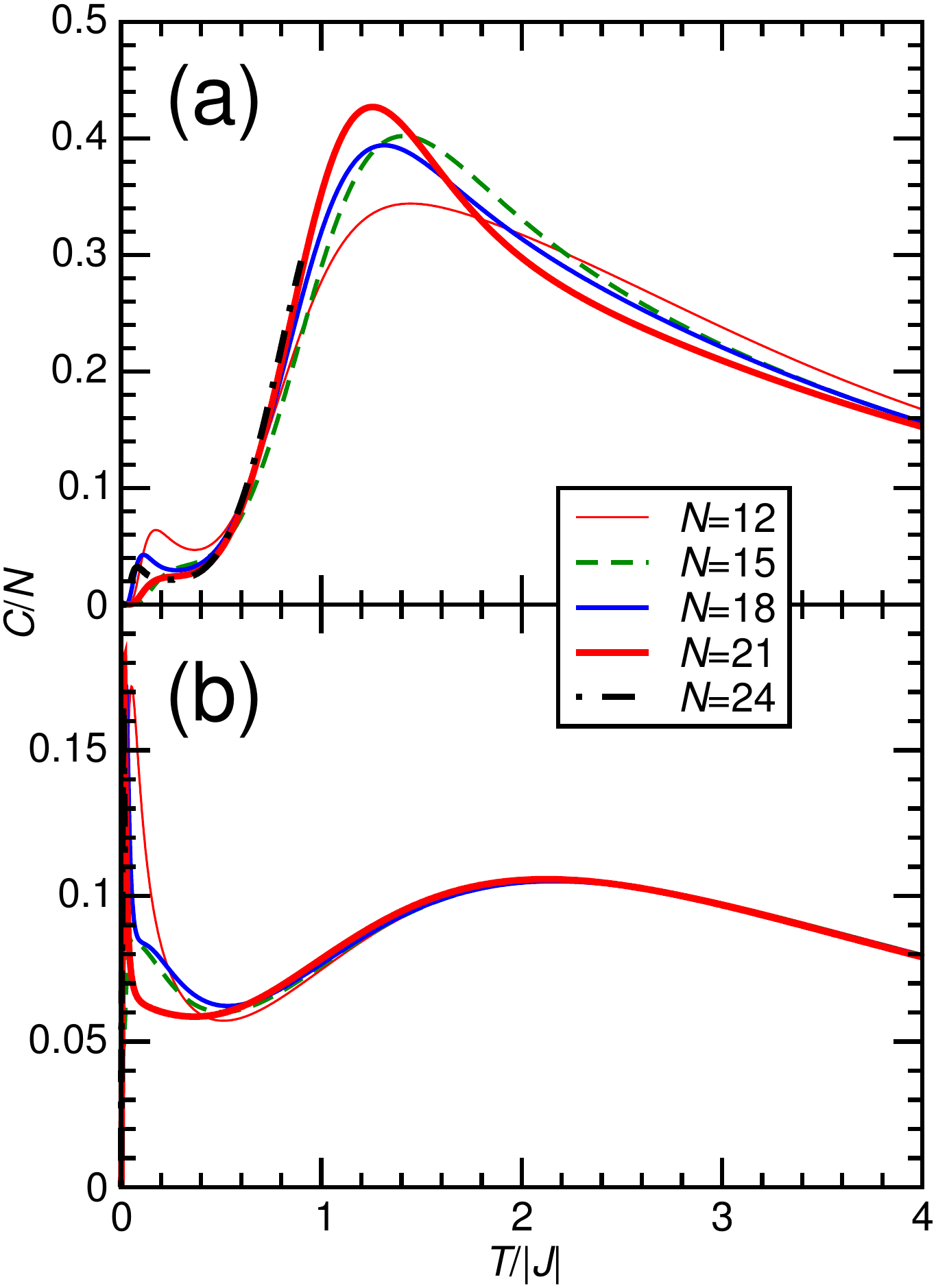}
\end{center}
\caption{(Color online) Specific heat per site
for the $S=1/2$ model with (a) $J < 0$ and (b) $J>0$.}
\label{figCall}
\end{figure}

The specific heat $C$ can be expressed in terms of the
eigenvalues of the Hamiltonian. Since we have all eigenvalues
for $N \le 21$, it is straightforward to obtain
the specific heat for all temperatures and both signs of $J$.
Fig.~\ref{figCall} shows the results of the specific heat per
site $C/N$. The case $J<0$ is shown in panel (a).
There is a finite-size maximum slightly above $T \approx \abs{J}$.
The large finite-size effects which are still observed here
are consistent with a phase transition around $T \approx \abs{J}$
in which case $C$ should become non-analytic for $N \to \infty$.
Because of a possible phase transition, we have tried to obtain
a low-temperature approximation to the specific heat for
$N > 21$, $J < 0$ by keeping low-energy states.
However, for $N=24$ even 12~462 low-lying states going up to energies as
high as $\Delta E \lesssim 12.6 \, \abs{J}$ turned out
to yield a specific heat which has sufficiently small
truncation errors only for temperatures $T \lesssim 0.9 \, \abs{J}$.
This result for $N=24$ (also included in Fig.~\ref{figCall}(a))
clearly does not include the maximum of the specific heat $C$.

Now we turn to the case $J > 0$ which is shown in panel (b) of
Fig.~\ref{figCall}. In this case,
finite-size convergence at high temperatures
is much faster than for $J < 0$.
This fast convergence indicates that there is no
phase transition associated to the high-temperature maximum
(the position of this maximum is at $T \approx 2.1225 \, J$
with a value $C \approx 0.105717 \, N$ for $N=21$). The reduced
finite-size effects and the smaller value of $C$ at the maximum
reflect that there is a substantially smaller energy
scale for $J > 0$ as compared to $J < 0$.

\begin{figure}[tb!]
\begin{center}
\includegraphics[width=\columnwidth]{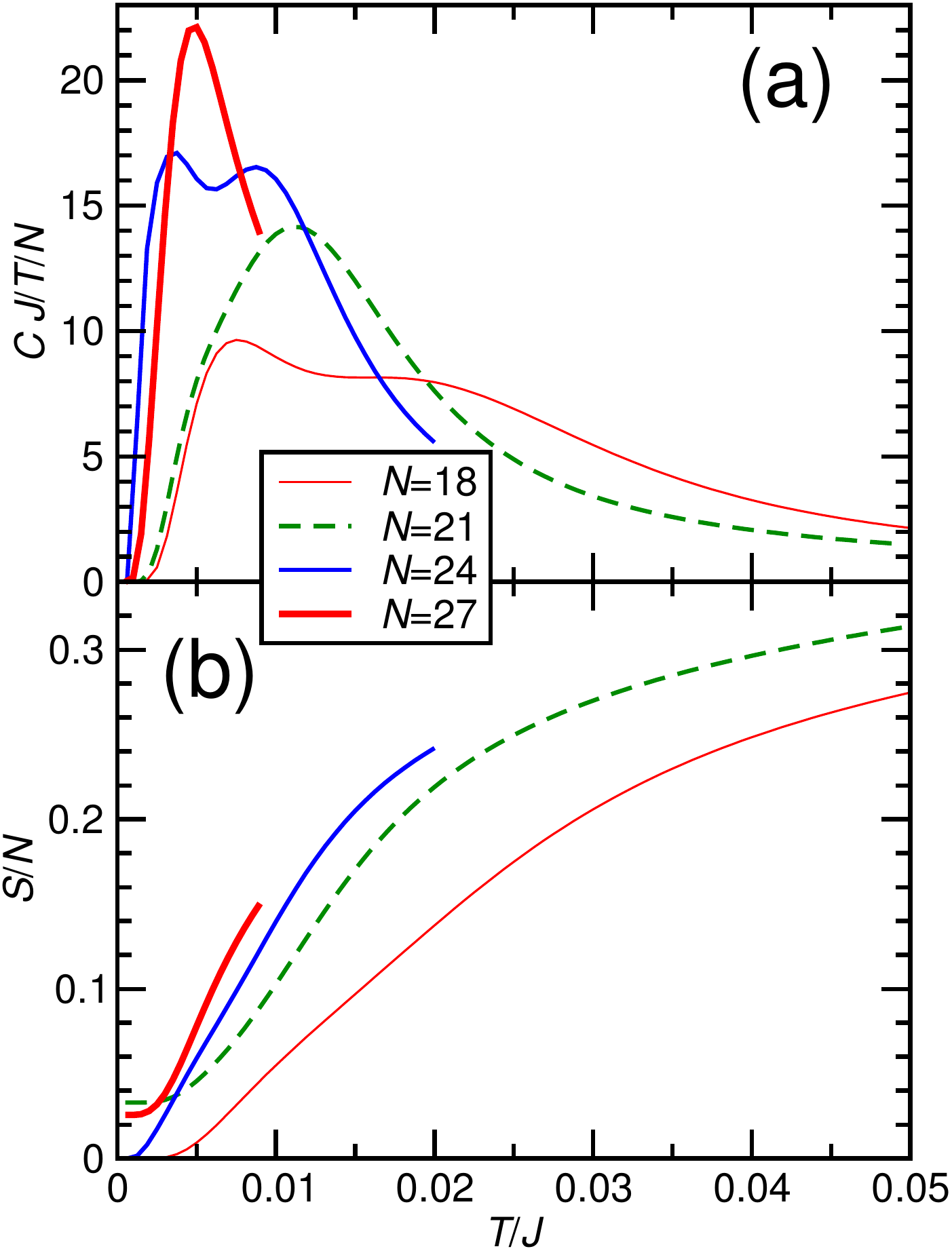}
\end{center}
\caption{(Color online) Low-temperature behavior of the
specific heat divided by temperature
$C/T$ (a) and entropy $S$ (b) per site $N$ for $J>0$.}
\label{figCSlt}
\end{figure}

For $J > 0$, a second peak emerges in the specific heat at
low temperatures, see Fig.~\ref{figCall}(b). In order to
investigate this in more detail, we use again the low-temperature
approximation for the specific heat obtained from the
low-lying part of the spectrum. For $N=24$ and $27$, we
have used a total of $7~029$ and $3~906$ eigenvalues,
respectively (the $N=24$ data is included in Fig.~\ref{figCall}(b),
but it is difficult to see there since it is valid only at
very low temperatures). Fig.~\ref{figCSlt} shows the specific
heat divided by temperature (panel (a)) and the entropy per site (panel (b))
in the low-temperature region for $J > 0$ and
system sizes $N=18$, $21$, $24$, and $27$. Our result
for the specific heat with $N=21$ obtained from the full
spectrum agrees with a previous result for $N=21$ based on
approximately 2~000 low-lying states.\cite{DFEBSL05}
The finite $T=0$ limit of the entropy for $N=21$ and $27$
in Fig.~\ref{figCSlt}(b) corresponds to the two-fold degeneracy
of the ground state for these system sizes, see Fig.~\ref{figSpec}(b).
Although the maximum value of $C/T$ increases with increasing $N$,
there are non-systematic finite-size corrections to the position
of this maximum. Thus, we can only conclude that if there is
a finite-temperature ordering transition for $J>0$, it should
have a very low transition temperature $T_c \lesssim J/100$.

Fig.~\ref{figCSlt}(b) shows that there is a remarkably large
entropy $S/N \approx 0.2\ldots 0.3$ associated to the
finite-size low-temperature peak of the specific heat.
This is comparable to the entropy associated to the
degeneracy of the classical ground states, see
section \ref{secEnumCl} below.
Therefore, this observation lends further support to
the interpretation\cite{DFEBSL05} of the low-energy states for
$S=1/2$ in terms of the classical ground states for $J > 0$.

\section{Classical System: lowest-energy configurations and spin-wave analysis}

\label{secCl}

We will now proceed with a discussion of the low-energy, low-temperature
properties of
the classical variant of the model (\ref{eqH}), {\it i.e.}, we will
assume that the $\vec{S}_i$ are unit vectors.
We will parametrize the spin at site $i$ by angles $\alpha_i$ and $\gamma_i$:
\begin{equation}
\vec{S}_i = \pmatrix{\cos \gamma_i \cos \alpha_i \cr \cos \gamma_i \sin \alpha_i \cr \sin \gamma_i } \, .
\label{eqSclXY}
\end{equation}
Because the $z$-components do not
contribute to the energy, configurations with extremal energy should
have spins lying in the $x$-$y$-plane ($\gamma_i =0$).
The energy $E(\{\alpha_i\})$ for a given
set of angles $\{ \alpha_i \}$, $\gamma_i =0$ is obtained from (\ref{eqH})
by identifying
\begin{eqnarray}
T^A_i &=& 2 \, \cos \left(\alpha_i\right) \, , \nonumber \\
T^B_i &=& 2 \, \cos\left(\alpha_i+ \Omega \right) \, , \label{eqTcl} \\
T^C_i &=& 2 \, \cos\left(\alpha_i- \Omega \right) \nonumber
\end{eqnarray}
with $\Omega = 2\, \pi/3$.

We will further be interested in small fluctuations
$\{ \alpha_i + \epsilon_i \}$,  $\{ \gamma_i = \tilde{\epsilon}_i \}$ around a ground-state configuration
$\{ \alpha_i , \gamma_i = 0\}$. The energy can then be expanded as

\begin{eqnarray}
E\left(\{\alpha_i + \epsilon_i \}\right)
&=& E\left(\{\alpha_i\}\right) + \sum_{i,j} \epsilon_i \, M_{i,j} \, \epsilon_j
 \nonumber \\
&+& E_{zz} + {\cal O}\left(\{ \epsilon_i , \tilde{\epsilon_j} \}^3\right) \, .
\label{eqDefM}
\end{eqnarray}
Here, $E_{zz}$ is a diagonal quadratic function of the out-of-plane
fluctuations $\tilde{\epsilon}_i$ which, to quadratic order,
decouples from the relevant degrees of freedom $\epsilon_i$. 
The eigenvalues $\sw_i$ of the symmetric matrix $M_{i,j}$
correspond to the spin-wave modes. The fact that $\{\alpha_i\}$ describes
a ground state implies $\sw_i \ge 0$. We will call a mode with
$\sw_i = 0$ `pseudo-Goldstone mode'.

\subsection{Ground states with small unit cell for $J < 0$}

\label{secCLgsNeg}

\begin{figure}[tb!]
\begin{center}
\includegraphics[width=0.9\columnwidth]{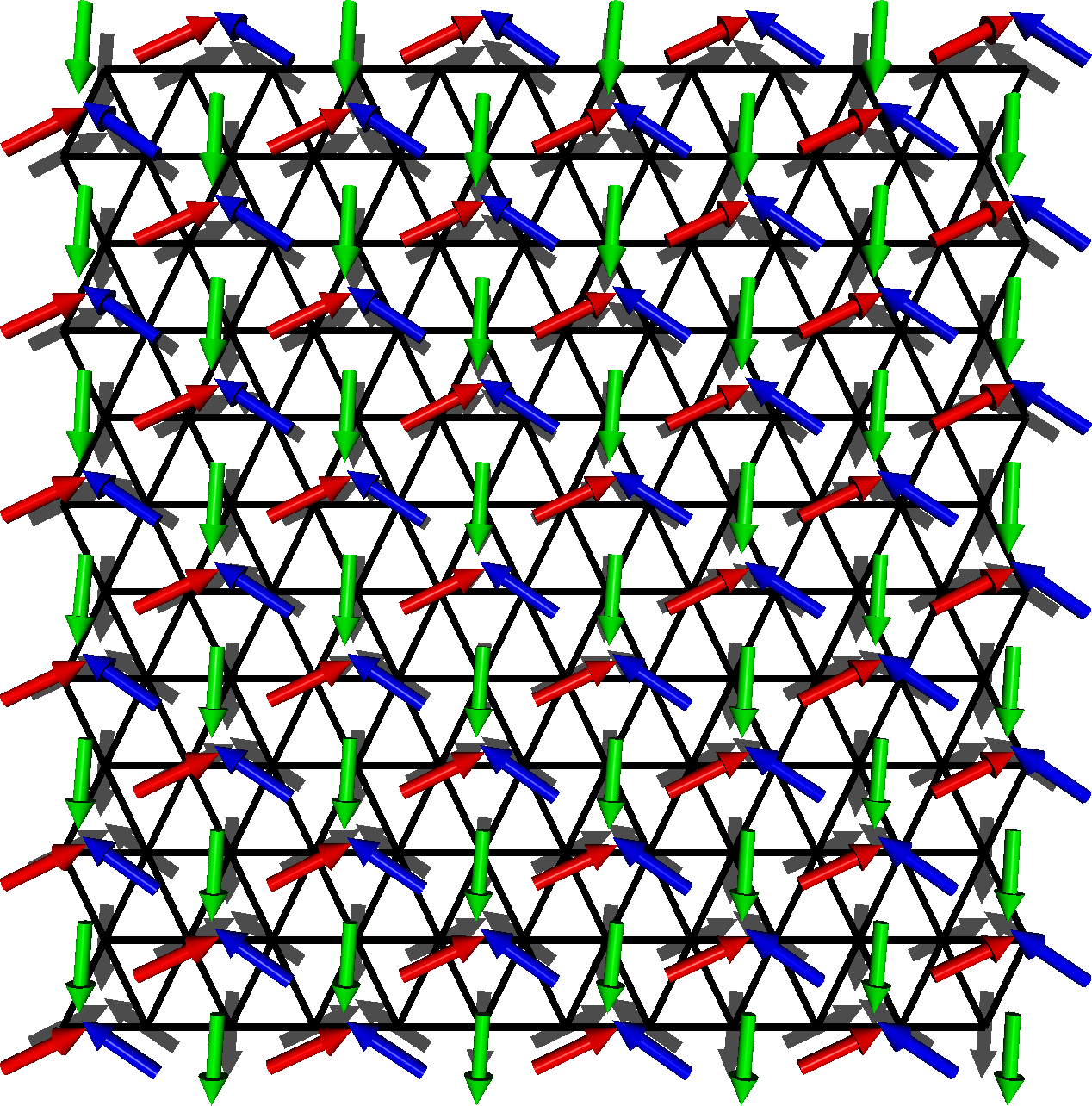}
\end{center}
\caption{(Color online) Snapshot of a configuration during a simulation
for $J<0$
at $T = 10^{-3} \,\abs{J}$ on a $12\times 12$ lattice. Periodic boundary
conditions are imposed at the edges.
}
\label{figSnapshotNeg}
\end{figure}

Let us first consider the case $J < 0$. Then a ground state is given
by a certain three-sublattice configuration where the angles
$\alpha_i$ between different sublattices differ by multiples of $2
\, \pi/3$.\cite{DEHFSL05,DFEBSL05} Fig.\ \ref{figSnapshotNeg} shows
such a low-temperature configuration as a snapshot which was taken 
during a Monte-Carlo simulation (details to be given in section \ref{secMC}
below).
The energy of such states
$E^{\Neg}_{\rm class.} = 6 \, J \, N$ is invariant under global
rotations of the spin configuration in the $x$-$y$-plane, {\it
i.e.}, there is a one-parameter family of ground states (note that
this invariance under a continuous group is not a symmetry of the
Hamiltonian).
We
parametrize this global rotational degree of freedom by an angle
$\alpha$ of the spins on one sublattice. Using a three-site unit
cell, we can exploit invariance of this ground state under
translations to represent the matrix (\ref{eqDefM}) in Fourier space
by the following $3 \times 3$ matrix

\begin{equation}
M = J\, \pmatrix{
6 & -2 \, {\cal A}  & -2 \, {\cal B} \cr
-2 \, {\cal A}^\star & 6 &  -2 \, {\cal C} \cr
-2 \, {\cal B}^\star & -2 \, {\cal C}^\star & 6 }
\label{eqM3Jneg}
\end{equation}
where
\begin{eqnarray}
{\cal A} &=&
{\rm e}^{i \, k_1} \, \sin^2\left(\alpha\right)
+ {\rm e}^{i \, k_2} \, \sin^2\left(\alpha+\Omega\right) \nonumber \\
&& + {\rm e}^{i \, k_3} \, \sin^2\left(\alpha-\Omega\right) \nonumber \, , \\
{\cal B} &=&
{\rm e}^{-i \, k_1} \, \sin^2\left(\alpha-\Omega\right)
+ {\rm e}^{-i \, k_2} \, \sin^2\left(\alpha\right) \nonumber \\
&& + {\rm e}^{-i \, k_3} \, \sin^2\left(\alpha+\Omega\right) \, ,
\label{eqDefABCneg} \\
{\cal C} &=&
{\rm e}^{i \, k_1} \, \sin^2\left(\alpha+\Omega\right)
+ {\rm e}^{i \, k_2} \, \sin^2\left(\alpha-\Omega\right) \nonumber \\
&& + {\rm e}^{i \, k_3} \, \sin^2\left(\alpha\right) \nonumber \, ,
\end{eqnarray}
and
\begin{equation}
k_1 = k_x \, , \quad
k_2 = -{k_x \over 2} + {\sqrt{3} \over 2} \, k_y \, , \quad
k_3 = -{k_x \over 2} - {\sqrt{3} \over 2} \, k_y \, .
\label{eqKdef}
\end{equation}

Let us analyze now the effect of the fluctuations by computing the
free energy associated with (\ref{eqDefM}). To this end we can
compute the partition function
%
%
\begin{equation}
Z_{\alpha} = {\rm e}^{-\beta H_0} ~Z_{zz}~\int \prod_{\vec{k}} d
\epsilon(\vec{k})~ {\rm e}^{-\beta \sum_{i,\vec{k}} \sw_i^{\alpha} (\vec{k})
\, \epsilon(\vec{k})^2} \, ,
\label{zsadpoint}
\end{equation}
where $\sw_i^{\alpha}(\vec{k})$ are the eigenvalues of (\ref{eqM3Jneg})
and $Z_{zz}$ is the Gaussian integral over the $N$ quadratic variables
corresponding to the out-of-plane fluctuations. 

Performing the Gaussian integral we get
\begin{equation}
Z_{\alpha} \sim {\rm e}^{-\beta
H_0}~Z_{zz}~ \prod_{i,\vec{k}} { \sqrt{\pi} \over \sqrt{  \beta \,
\sw_i^{\alpha}(\vec{k})   }} \, ,
\end{equation}
which yields the free energy as
\begin{equation}
F = H_{0} + F_{zz}+ {N\, \ln\beta \over 2 \, \beta}
 + {1 \over 2 \, \beta} \sum_{i,\vec{k}} \ln(\sw_i^{\alpha}(\vec{k}))
 + \ldots \, ,
\label{freeEn}
\end{equation}
%
where $F_{zz} = -\ln Z_{zz}/\beta$.

\begin{figure}[tb!]
\begin{center}
\includegraphics[width=\columnwidth]{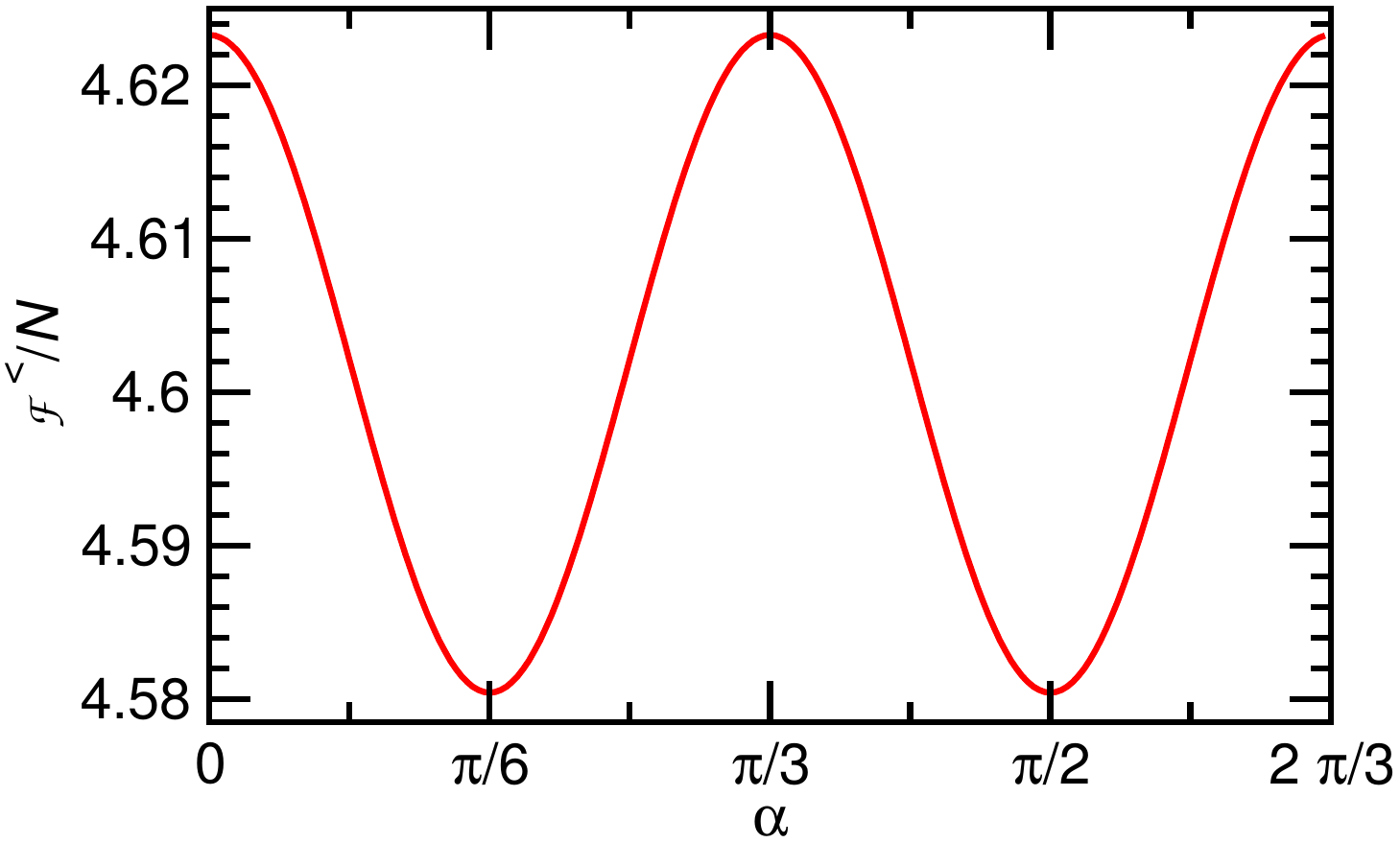}
\end{center}
\caption{(Color online) Low-temperature contribution to the
free energy (\ref{FlucJneg})
for $J<0$ of the fluctuations above the $120^\circ$ ground state
as a function of the spin angle $\alpha$.}
\label{figFlucJneg}
\end{figure}

\begin{figure}[tb!]
\begin{center}
\includegraphics[width=\columnwidth]{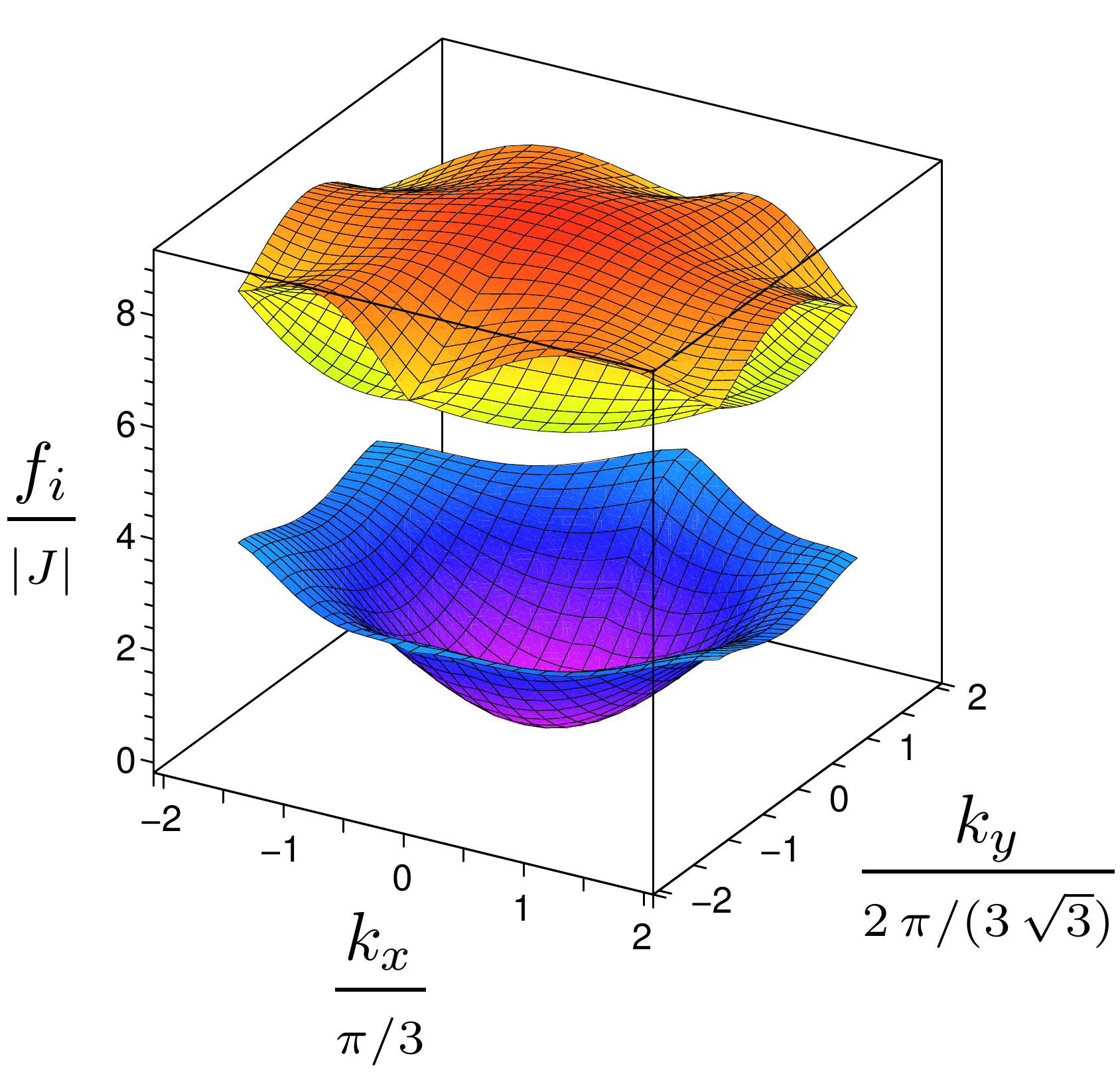}
\end{center}
\caption{(Color online) The three eigenmodes $\sw_i$ for the $120^\circ$
ground state with $J<0$, assuming a lock-in of the ground state at
$\alpha = (2\,n+1)\,\pi/6$. Note that the shaded surfaces extend only
over the first Brillouin zone.}
\label{figClSpecJneg}
\end{figure}

The low-temperature behavior is therefore determined by the
following contribution of the fluctuations to the free energy
\begin{equation}
{\cal F}^{\Neg}(\alpha) =
\sum_{i,\vec{k}} \ln
\sw_i^\alpha(\vec{k}) = \sum_{\vec{k}} \ln \det M \, ,
\label{FlucJneg}
\end{equation}
where $M$ is the matrix (\ref{eqM3Jneg}). The result of the
integral (\ref{FlucJneg}) is shown
in Fig.~\ref{figFlucJneg}. ${\cal F}^{\Neg}(\alpha)$ is
a $2 \, \pi/3$-periodic function since the spin angles on the
different sublattices differ by $2 \, \pi/3$. Hence, it is
sufficient to consider $\alpha \in [0, 2 \, \pi/3]$. One
observes that ${\cal F}^{\Neg}$ and thus the low-temperature
limit (\ref{freeEn}) of the free energy has minima at
$\alpha = (2\,n+1)\,\pi/6$, $n=0,1,\ldots,5$. This implies
that the $120^\circ$ classical ground-state configuration locks
in at these angles for $T \to 0$.
This lock-in can indeed be verified in histograms of Monte-Carlo
simulations, see Ref.~\onlinecite{hfm2006} for a planar variant of
this model and also Fig.~\ref{figStatPhi} below.
Lock-in of the classical ground-state configuration at $\alpha = (2\,n+1)\,\pi/6$ 
follows also from the semiclassical approach.\cite{DFEBSL05}
In this approach the ground-state energy is given by 
\begin{equation}
E^<_{\rm semclass.}(\alpha) = E^<_{\rm class.} + 6\,J\,S
 + \frac{1}{2\,N}\sum_{\vec{k} \atop i=1,2,3} \omega^<_i(\vec{k},\alpha)\,,
\label{esemclass1}
\end{equation}
where $\omega^<_i(\vec{k},\alpha)$ are the three sheets of spin-wave (SW) frequencies 
obtained from the linear Holstein-Primakoff approximation and where the sum over 
$\vec{k}$ runs over the magnetic Brillouin zone. The SW frequencies are connected 
with the classical eigenmodes $f^<_{{\rm class.},i}$ by 
$ f^<_{{\rm class.},i}(\alpha) =  (\omega^<_i(\vec{k},\alpha))^2/(24 \,S^2\,|J|)$. One finds that 
$E^<_{\rm semclass.}(\alpha)$ has minima at $\alpha = (2\,n+1)\,\pi/6$.
The expressions for 
$f^<_{{\rm class.},i}(\alpha)$ are too cumbersome to be given explicitly, instead 
Fig.~\ref{figClSpecJneg} shows a plot of the three eigenfrequencies $\sw_i$ at 
$\alpha = (2\,n+1)\,\pi/6$. The lowest {sheet} has a unique quadratic minimum at $\vec{k}=0$ with
$\sw^{\Neg}_{\rm class.,1}(\vec{k}) \approx 9\, \abs{J} \,
\vec{k}^2/8$ for $\vec{k} \approx 0$.

\subsection{Ground states with small unit cell for $J > 0$}

\label{secCLgsPos}

\begin{figure}[tb!]
\begin{center}
\begin{center}
\includegraphics[width=0.9\columnwidth]{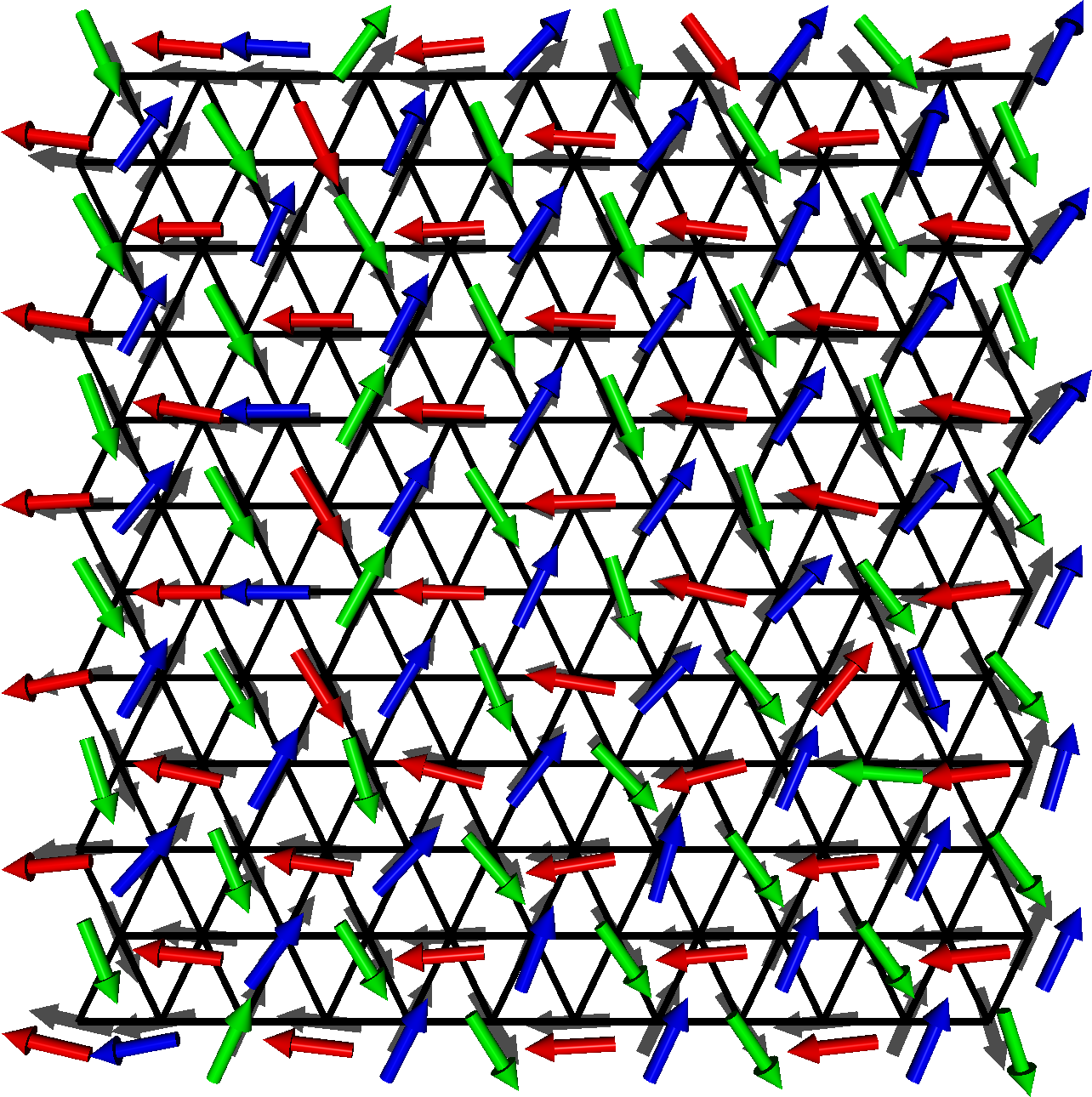}
\end{center}
\end{center}
\caption{(Color online) Snapshot of a low-temperature configuration 
during a simulation for $J>0$
at $T = 10^{-3} \,J$ on a $12\times 12$ lattice. Periodic boundary
conditions are imposed at the edges.
Different colors are used for each of the three sublattices.}
\label{figSnapshotPos}
\end{figure}

Now we turn to the case $J > 0$. There is a first ground
state\cite{DEHFSL05,DFEBSL05} described by $\alpha_i = \alpha$ with
an arbitrary angle $\alpha$. This `ferromagnetic' state has energy
$E^{\rm ferro}_{\rm class.} = -3 \, J \, N$. However, for $J > 0$
there is another ground state with a small unit
cell,\cite{DEHFSL05,DFEBSL05} again with three sublattices where the
angles $\alpha_i$ between different sublattices differ by multiples
of $2 \, \pi/3$. Also the energy $E^{\Pos}_{\rm class.} = -3 \,
J \, N$ is invariant under global rotations. The latter state
is illustrated by the global structure of Fig.\ \ref{figSnapshotPos}
which shows a snapshot of a low-temperature configuration taken during
a Monte-Carlo simulation (details again to be given in section \ref{secMC} below).
Note that the sense of orientation around a triangle, {\it i.e.}, the chirality
of the spins in Fig.\ \ref{figSnapshotPos} is exactly the opposite
of Fig.\ \ref{figSnapshotNeg}.

The ferromagnetic state
is the simplest case for the computation of fluctuations since
$M_{i,j}$ is diagonalized by a Fourier transformation. One finds the
modes
\begin{eqnarray}
{\sw^{\rm ferro}_{\rm class.}(k_x, k_y) \over 4 \, J} =
{3 \over 4}
&+& \sin\left(\alpha\right)\sin\left(\alpha-\Omega \right)
\cos\left(k_1\right) \nonumber \\
&+& \sin\left(\alpha+\Omega\right)\sin\left(\alpha-\Omega \right)
\cos\left(k_2\right) \nonumber \\
&+& \sin\left(\alpha\right)\sin\left(\alpha+\Omega \right)
\cos\left(k_3\right) \, ,
\label{eqSWferro}
\end{eqnarray}
with the $k_i$ defined in Eq.~(\ref{eqKdef}).
As for the case $J < 0$, the classical frequencies $\sw^{\rm ferro}_{\rm class.}$
are proportional to the squares of the SW frequencies $\omega^{\rm ferro}$
obtained from a linear Holstein-Primakoff approximation:\cite{DFEBSL05}
$\sw^{\rm ferro}_{\rm class.} =  \left(\omega^{\rm ferro}\right)^2
/ \left(12 \, S^2 \, J\right)$.

By computing the contribution of the modes $\sw^{\rm ferro}_{\rm class.}$
to the free
energy we find minima for $\alpha= n\,\pi/3$, $n=0,1,2,\ldots$, so
that the spins in the ferromagnetic structure lock in to the
lattice directions of the triangular lattice. For the lock-in values
of $\alpha$, $\sw^{\rm ferro}_{\rm class.}$ depends only on one
of the $k_i$ and has a line of zeros in the perpendicular
direction in momentum space.

The three-sublattice state leads to the following $3 \times 3$
matrix in Fourier space:
\begin{equation}
M = J\, \pmatrix{
3 & 2 \, \tilde{\cal A} & 2 \, \tilde{\cal B} \cr
2 \, \tilde{\cal A}^\star & 3 &  2 \, \tilde{\cal C} \cr
2 \, \tilde{\cal B}^\star &  2 \, \tilde{\cal C}^\star & 3 } \, ,
\label{eqM3Jpos}
\end{equation}
where
\begin{eqnarray}
\tilde{\cal A} &=&  {\rm e}^{i \, k_2} \, \sin(\alpha +\Omega) \, \sin(\alpha -\Omega)
\nonumber \\
 &+& \left( {\rm e}^{i \, k_1} \, \sin(\alpha + \Omega)
 + {\rm e}^{i \, k_3} \, \sin(\alpha - \Omega) \right) \, \sin(\alpha) \, , \nonumber \\
\tilde{\cal B} &=& {\rm e}^{-i \, k_1} \, \sin(\alpha +\Omega) \, \sin(\alpha -\Omega)
\nonumber \\
 &+& \left({\rm e}^{-i \, k_3} \, \sin(\alpha + \Omega)
 + {\rm e}^{-i \, k_2} \, \sin(\alpha - \Omega) \right) \, \sin(\alpha) \, , \nonumber  \\
\tilde{\cal C} &=&  {\rm e}^{i \, k_3} \, \sin(\alpha +\Omega) \, \sin(\alpha -\Omega)
\label{eqDefABCpos} \\
 &+& \left( {\rm e}^{i \, k_2} \, \sin(\alpha + \Omega)
 + {\rm e}^{i \, k_1} \, \sin(\alpha - \Omega) \right) \, \sin(\alpha) \, . \nonumber
\end{eqnarray}
For $\alpha = n\,\pi/3$, diagonalization of (\ref{eqM3Jpos})
leads to three completely flat branches
\begin{equation}
\sw^{\Pos}_{\rm class.,1}(\vec{k}) = 0 \, , \quad
\sw^{\Pos}_{\rm class.,2}(\vec{k}) =
\sw^{\Pos}_{\rm class.,3}(\vec{k}) = {9 \over 2} \, J \, .
\label{eqJposSW}
\end{equation}
In particular the lowest branch $\sw^{\Pos}_{\rm class.,1} = 0$
corresponds to a branch of soft modes.
In real space these soft modes correspond to the rigid rotation of
one single triangle.\cite{DFEBSL05}
Note that there is no such flat branch of soft modes
for a value of $\alpha$ which is not an integer multiple of $\pi/3$.

When computing the contribution of fluctuations around these
configurations ($\alpha = n\,\pi/3$) to the free energy, one finds
that one third of the modes are quartic instead of quadratic.
This yields a free energy of the form:
\begin{equation}
F = H_{0} +F_{zz} + F_{xy} \, ,
\end{equation}
where, again, $F_{zz} \sim  {N\, \ln\beta / (2\, \beta)}$
corresponds to the trivial contribution of out-of-plane fluctuations and
(compare also Ref.\ \onlinecite{CHS92})
\begin{equation} F_{xy} =  {N\, \ln\beta \over \, 3 \, \beta}
 + {N \, \ln\beta \over 12 \, \beta} + \ldots
\label{freesoft}
\end{equation}
At low temperatures, this term dominates the free energy.
The flat branch of soft modes reduces the coefficient of
$\ln\beta/\beta$ from $N/2$ as in the case of only
quadratic modes (compare (\ref{freeEn})) to $N/3+N/12 = 5\,N/12$.
This implies two things: firstly, the angles of
the $120^\circ$ state should lock in at $\alpha = n\,\pi/3$
for low temperatures. Secondly, a thermal
order-by-disorder mechanism should favor the
$120^\circ$ state over the ferromagnetic state
for $T \to 0$. 

As in the case of the ferromagnetic state one finds that the relation 
$\sw^{\Pos}_{{\rm class.},i} =  \left(\omega^{>}_i\right)^2
/ \left(12 \, S^2 \, J\right)$, where $\omega^{>}_i$, $i=1,\,2,\,3,$ are
the SW frequencies obtained from a linear Holstein-Primakoff
approximation,\cite{DFEBSL05,DEHFSL05} holds for arbitrary values of
$\alpha$. Using the results for $\omega^{\rm ferro}(\vec{k},\alpha)$ and  
$\omega^{>}_i(\vec{k},\alpha)$ to calculate 
semiclassical ground-state energies $E^{\rm ferro}_{\rm semclass.}(\alpha)$ 
and  $E^{\Pos}_{\rm semclass.}(\alpha)$ in the same manner as in (\ref{esemclass1})    
one finds that both are minimal at $\alpha = n\,\pi/3$ and that  
$E^{\Pos}_{\rm semclass.}(n\,\pi/3) < E^{\rm ferro}_{\rm semclass.}( n\,\pi/3)$. Thus 
the semiclassical approach is fully consistent with the classical findings.

\subsection{Enumeration of ground states for small $N$}

\label{secEnumCl}

\begin{table}
\caption{\label{tab:gsdeg}
Number ${\cal D}_N$ of classical ground states for $J > 0$ on a lattice
of size $N$ with one angle fixed. The $n_g$ in the decomposition
${\cal D}_N = \sum_{g} n_g$ denote the number of ground states
with $g$ pseudo-Goldstone modes.
}
\begin{ruledtabular}
\begin{tabular}{lll}
$N$ & ${\cal D}_N = \sum_{g} n_g$ & ${\ln {\cal D}_N \over N}$ \\ \hline
$12$ & $40 = 6_1 + 31_2 + 2_3 + 1_4$ & $0.3074$ \\
$15$ & $102 = 60_1 + 20_2 + 20_3 + 2_5$ & $0.3083$ \\
$18$ & $286 = 92_1 + 112_2 + 51_3 + 30_4 + 1_6$ & $0.3142$ \\
$21$ & $688 = 260_1 + 210_2 + 203_3 + 14_5 + 1_7$ & $0.3111$ \\
$24$ & $1838 = 384_1 + 958_2 + 199_3 + 280_4 + 16_6 + 1_8 $ & $0.3132$ \\
$27$ & $5054 = 1068_1 + 972_2 + 2257_3 + 351_4 + 378_5$ & $0.3158$ \\
 & $\phantom{5054 =} + 9_6 + 18_7 + 1_9$ &  \\
\end{tabular}
\end{ruledtabular}
\end{table}

Direct computer inspection of all states with angles $\alpha_i = n_i \, \pi/3$,
$n_i \in \{0, 1, 2, 3, 4, 5\}$ for $N=12$ shows\cite{DEHFSL05,DFEBSL05}
that there are further ground states for $J > 0$.
On the one hand, CPU time for a
similar enumeration of all $6^N$ such configurations becomes prohibitively
big for $N \gtrsim 15$. On the other hand, all known ground states turn
out to have mutual angles which are multiples of $2 \, \pi/3$.
Furthermore, we eliminate a global rotational degree of freedom by
fixing one arbitrary angle $\alpha_0 = \pi$. Then there remain just
$3^{N-1}$ configurations with $\alpha_i \in \{\pi/3, \pi, -\pi/3\}$
to be enumerated. Direct enumeration of these $3^{N-1}$ configurations
can be carried out with reasonable CPU time for $N \le 27$, but
becomes quickly impossible for larger $N$.
We have therefore performed such enumerations for $N \le 27$, using the
same lattices as in section \ref{secED}.

The number of ground states ${\cal D}_N$ determined in this manner
is given in Table~\ref{tab:gsdeg} for $J > 0$.
Note that the ordered states which we have described
in section \ref{secCLgsPos} are
just two of the ${\cal D}_N$ states, but there are many further
ground states which can be interpreted as defects in and domain walls
between the ordered states.\cite{DEHFSL05,DFEBSL05}
Indeed, also closer inspection of the snapshot
shown in Fig.\ \ref{figSnapshotPos} reveals the presence
of defects in the three-sublattice structure.
The $240 = 6 \, {\cal D}_{12}$
states described previously\cite{DEHFSL05,DFEBSL05} for $N=12$ are
recovered by a global rotation of the angles such that $\alpha_0$
takes on the six values $\alpha_0 = n \, \pi/3$.
The last column of Table~\ref{tab:gsdeg} lists
$\ln\left({\cal D}_N\right)/N$. The fact that these numbers stay
almost constant indicates a finite ground-state entropy per site
slightly above $0.3$ in the thermodynamic limit.

It is straightforward to derive the $N \times N$ matrix $M_{i,j}$ defined in
(\ref{eqDefM}) for any ground state and diagonalize it.
Among the eigenmodes $\sw_i$, one
can then identify the $g$ pseudo-Goldstone modes $\sw_i = 0$ and in turn count
the number $n_g$ of ground states with $g$ pseudo-Goldstone modes. These numbers
are also given in Table~\ref{tab:gsdeg} in the form ${\cal D}_N = \sum_{g} n_g$.
One observes that all ground states have at least one pseudo-Goldstone mode.
There are at most $N/3$ pseudo-Goldstone modes, and there is only one ground
state with this maximal number of pseudo-Goldstone modes
corresponding to the three-sublattice state described in
section \ref{secCLgsPos} (apart from $N=15$; however, this lattice
is special in that it has a period three translational symmetry
$T_y^3 = {\mathbf 1}$).

There are $N/3$ ground states which differ from the perfect
three-sublattice state by a rigid rotation of the spins on certain triangles
by an angle $2\,\pi/3$, and another $N/3$ ground states
where the spins on a different set of triangles are rotated by $-2\,\pi/3$.
These ground states have two pseudo-Goldstone modes less
than the three-sublattice state. The data in Table~\ref{tab:gsdeg}
show that these $2\,N/3$ configurations with $N/3-2$ pseudo-Goldstone
modes account for all states with the second largest number of
pseudo-Goldstone modes for $N \ge 21$. This indicates that
ground states deviating from the homogeneous three-sublattice state
are obtained at the expense of pseudo-Goldstone modes.
At finite temperature such ``inhomogeneous'' ground states
are then penalized by an entropic cost because of the reduced number
of pseudo-Goldstone modes. This indicates that ground states with
bigger deviations from the three-sublattice ground state have
a higher free energy for small $T$ since they have less soft modes.
Thus, the global minimum of the free energy is the perfectly ordered
state with many close-by configurations which deviate only locally
from the perfect order. These arguments predict a thermal
order-by-disorder selection of the $120^\circ$ state
among the macroscopic number of ground states
for $T \to 0$ and $J > 0$.

The above enumeration procedure can also be performed for $J < 0$.
In fact, in this case we have carried it out twice, first
with $\alpha_i \in \{\pi/3, \pi, -\pi/3\}$ and $\alpha_0 = \pi$,
and then again with all $\alpha_i$ shifted by $\pi/6$ in order to
match the lock-in predicted in section \ref{secCLgsNeg}.
In sharp contrast to the large degeneracy found for $J > 0$, we confirm
that the ground state is unique (up to global rotations of all
angles) for $J < 0$ and thus identical to the three-sublattice state
described in section \ref{secCLgsNeg}. Diagonalization of the
corresponding $N \times N$ matrix $M$ yields one pseudo-Goldstone mode,
in complete agreement with the $\sw_i$ shown in
Fig.~\ref{figClSpecJneg} which have just one zero, namely
$\sw^{\Neg}_{\rm class.,1}(\vec{k}=0) = 0$.

\section{Classical system: Monte Carlo analysis}

\label{secMC}

Section \ref{secCl} already contained some discussion of the
low-temperature properties in the classical limit. The results
of section \ref{secED} lead us to suspect a finite-temperature
transition in the quantum system at least for $J < 0$. Indeed, a
finite-temperature phase transition is allowed \cite{MW1966}
since the model has only discrete and no continuous symmetries
(compare section \ref{sec:Hsym}). Since
such a finite-temperature phase transition should be a classical
phase transition, we may hope to gain insight into its universality
class by studying the classical counterpart of the model. This
motivates us to present results of a Monte-Carlo simulation of
the classical model, treating the spins $\vec{S}_i$ as classical
$O(3)$ vectors. Simulations were performed on square lattices
with diagonal bonds (topologically equivalent to the triangular
lattice) and periodic boundary conditions.

First, we have used a standard single-spin flip Metropolis algorithm
(see, e.g., Ref.~\onlinecite{LB00}). Some results obtained from such
simulations have already been published in Ref.~\onlinecite{hfm2006},
but the results to be presented here have been obtained from new runs
using the `Mersenne Twister' random number generator.\cite{P:matsu98}
In order to determine error bars,
we have used between 100 and 400 independent simulations for $J<0$.

For $J > 0$, the standard single-spin flip algorithm turns out
to be no longer ergodic for temperatures $T\lesssim 0.1\,J$.
Such problems are in fact
expected in view of the large ground-state degeneracy which we discussed in
section \ref{secEnumCl}. In this region we have
therefore used the parallel tempering Monte Carlo method (also known as
exchange Monte Carlo -- see Refs.\ \onlinecite{HN96,eM98,Hans97,KTHT06} and
references therein). In this framework, $n$ simulations are performed in
parallel, each at a different temperature using the standard Metropolis
algorithm. Periodically, the exchange between the configurations of two
simulations consecutive in temperature is proposed and accepted depending
on the energy balance of such a move. A careful choice of the temperature
points allows each configuration to shuffle through the entire temperature
range during the simulation, greatly reducing the probability of getting
stuck in a local minimum of the free energy. In principle, this allows an
efficient exploration of the phase space, while not having to wait for
rethermalization of the systems after each configuration switch.
Strategies to optimize the choice of the temperature grid have been
proposed (see, e.g., Ref.~\onlinecite{KTHT06}), but we simply opted for
constructing a fine-grained temperature set using the rule of thumb that
the probability for a configuration switch to be accepted should always be
at least around $70\%$ to $80\%$.  The resulting grid consists of
at least 96 points for
$T/J \in\left[0.01,0.7\right]$, where of course most points lie in the
low-temperature region. After an initial thermalization,
observables are sampled until convergence of their error bars is observed
(the typical duration of a simulation being at least $3\times10^7$
Monte-Carlo sweeps per system). We would like to insist that even if
parallel tempering is adequate to the task of studying the low-temperature
properties of such a highly degenerate frustrated magnet, it is still by
no means easy to obtain physically relevant data at such a low temperature
for continuous spherical spins as we shall see later.

\subsection{Specific heat}

\begin{figure}[tb!]
\begin{center}
\includegraphics[width=\columnwidth]{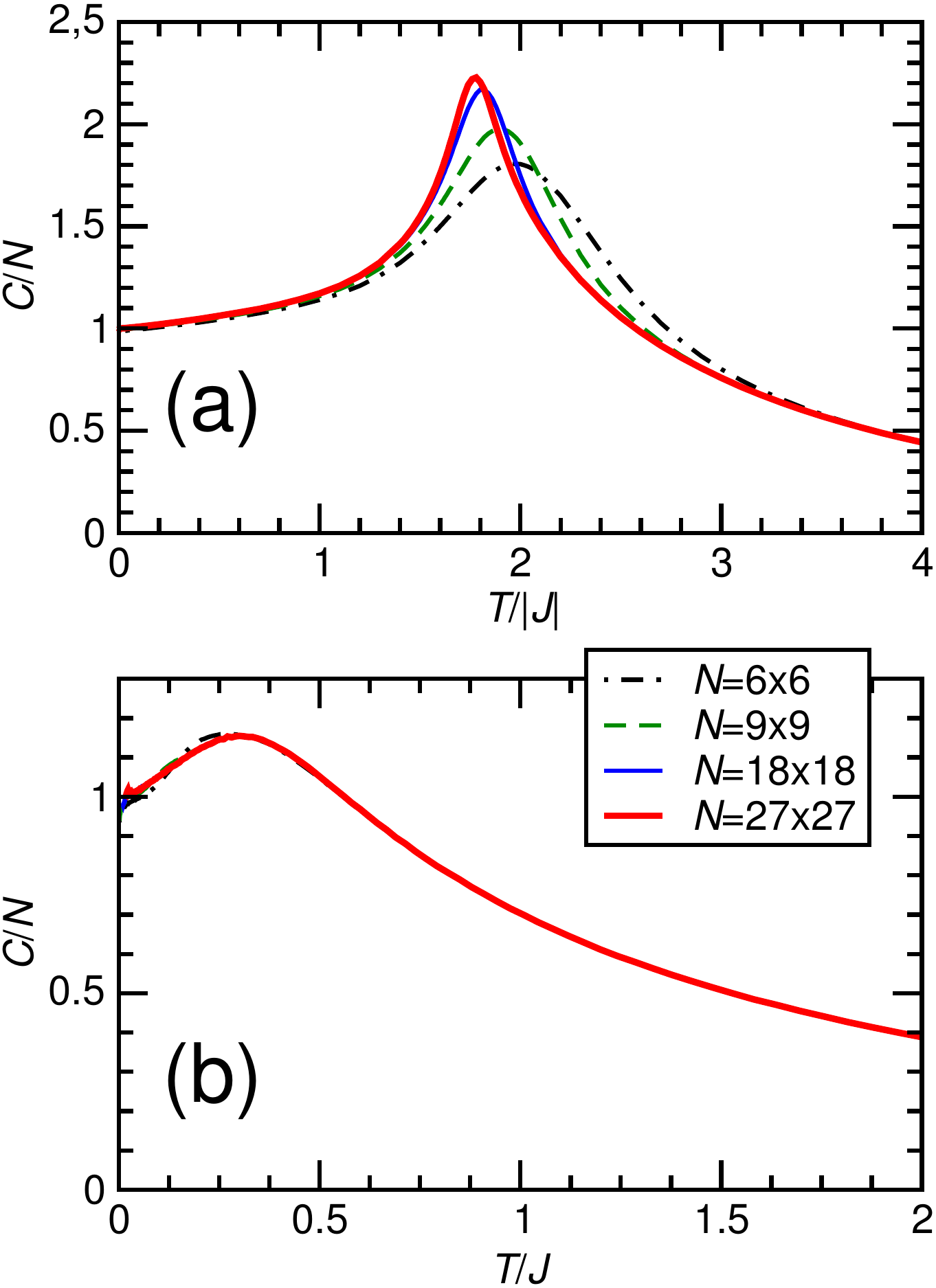}
\end{center}
\caption{(Color online) Specific heat per site
for the classical model with (a) $J < 0$ and (b) $J>0$.
Error bars do not exceed the width of the lines.}
\label{figCcl}
\end{figure}

\begin{figure}[tb!]
\begin{center}
\includegraphics[width=\columnwidth]{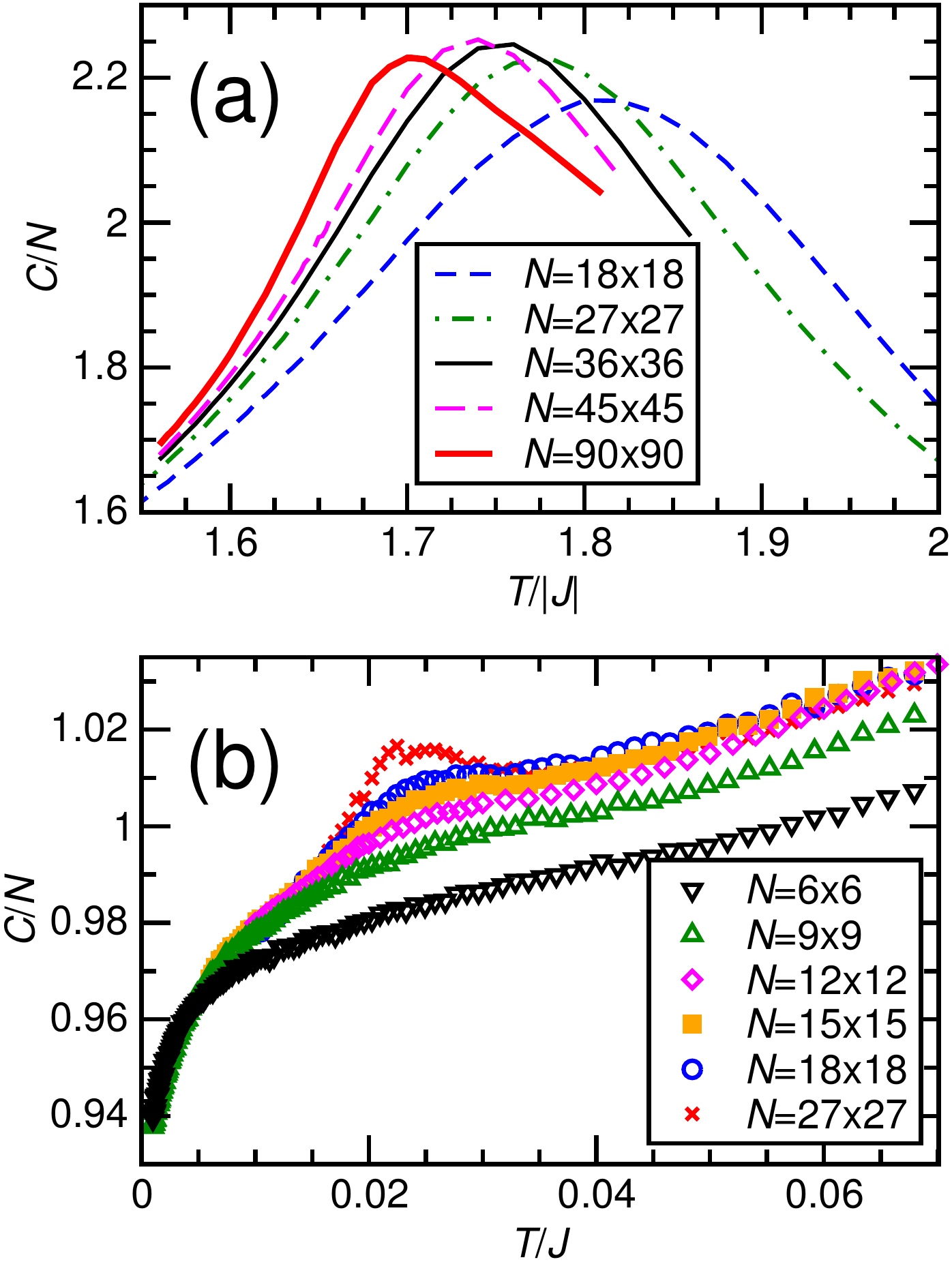}
\end{center}
\caption{(Color online) Zooms of the
specific heat per site for the classical model:
(a) in the vicinity of the maximum for $J<0$,
(b) in the low-temperature region for $J>0$.
Statistical errors do not exceed the width of the
lines or the size of the symbols, respectively.
}
\label{figCzoomed}
\end{figure}

Fig.~\ref{figCcl} shows results for the specific heat of the classical
model for $J<0$ (panel (a)) and $J>0$ (panel (b)). Statistical errors
should be at most on the order of the width of the lines.
Although all
lattice sizes are bigger than those used previously for the quantum model,
there are remarkable similarities of the specific heat of the classical
model shown in Fig.~\ref{figCcl}
with the specific heat of the quantum
model, see Fig.~\ref{figCall}. For $J<0$, a singularity seems to
develop in the
specific heat for temperatures around $T \approx 1.5 \, \abs{J}$,
signaling a phase transition. For $J>0$, there is also a broad maximum at
`high' temperatures $T \approx 0.3 \, J$. The finite-size effects for the
latter maximum are small indicating that this does not correspond to a phase
transition. In this case, the interesting features of the specific heat
lie in the low-temperature region, as displayed in
Fig.~\ref{figCzoomed}(b). As the system size increases, one can see that a
small peak builds up in the specific heat for $T \approx 0.02 \, J$. This
seems to indicate that  a phase transition might occur around that temperature,
two orders of magnitudes smaller than for $J<0$.  We are unfortunately not
on a par with the $J<0$ data, as the CPU requirements are too steep to
secure relevant data for systems larger than $27\times 27$ sites even
though the specific heat is a comparably robust quantity, and it
is clear that other observables are needed to conclude on the existence of
this phase transition.

An important difference between the $S=1/2$ and the classical model arises
at low temperatures: the specific heat of the quantum system has to vanish
upon approaching the zero temperature $\lim_{T \to 0} C/N = 0$, while due to
the remaining continuous degrees of freedom, the specific heat of the
classical system approaches a finite value for $T \to 0$. For $J<0$, the
equipartition theorem predicts an $N/2$ contribution to the specific heat
for each transverse degree of freedom which yields $\lim_{T \to 0} C/N =
1$, in excellent agreement with the results depicted in the panel (a)  of
Fig.\ \ref{figCcl}. For $J>0$, one must take into account the fact that
the flat soft-mode branch of the three-sublattice state is expected to
contribute only $N/12$ to the specific heat. Thus for $J>0$ one should
expect $\lim_{T \to 0} C/N = 11/12 = 0.916666...$. As can be seen in Fig.\
\ref{figCzoomed}(b), we observe a specific heat lower than one in the
low-temperature region along with a downward trend as $T$ goes to 0 for
all the system sizes studied.
However, according to the data which we have at our disposal, it seems
that one would have to go to very low temperatures $T < 10^{-3}\,J$ in
order to verify the prediction for the zero-temperature limit.

Returning to the finite-temperature transition for $J<0$,
Fig.\ \ref{figCzoomed}(a) shows a zoom into the relevant temperature
range, including data for up to $N=90\times90$ spins. At these bigger
system sizes, the position of the maximum continues to shift to lower
temperatures and the maximum sharpens. However, the $N=45\times45$
and $90\times90$ curves in Fig.\ \ref{figCzoomed}(a) demonstrate
that the maximum value of the specific
heat starts to decrease as one goes to system sizes beyond
$N=45\times45$. This implies that the exponent $\alpha$ which characterizes
the divergence of the specific heat at the critical temperature is very
small or maybe even negative.

\subsection{Sublattice order parameter, Binder cumulant, and
transition temperature for $J<0$}

According to subsection \ref{secCLgsNeg}, we expect that the phase
transition observed for $J<0$ is a transition into a three-sublattice
ordered state. This ordering is indeed
exhibited at least at a qualitative level by snapshots of Monte-Carlo
simulations at low temperatures (compare Fig.\ \ref{figSnapshotNeg}).
In addition, one observes in Fig.\ \ref{figSnapshotNeg}
that the spins are lying essentially in the $x-y$-plane for low temperatures.

\begin{figure}[tb!]
\begin{center}
\includegraphics[width=\columnwidth]{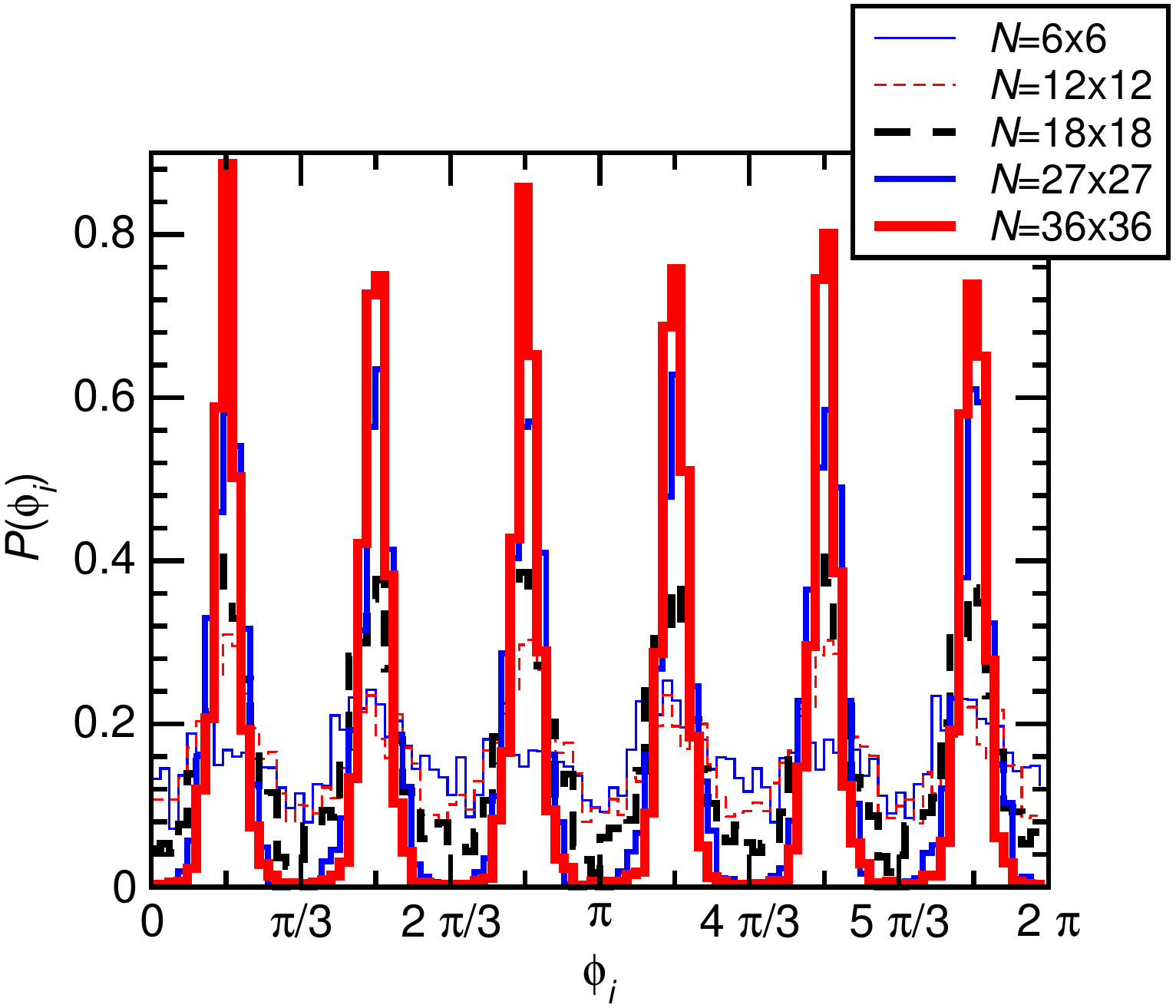}
\end{center}
\caption{(Color online)
Histogram of in-plane angles $\phi_i$ for $J<0$.
Averaging has been performed over
1000 independent configurations at $T = 10^{-3} \,\abs{J}$.
}
\label{figStatPhi}
\end{figure}

Furthermore, we expect a lock-in of the spins
to one of 6 symmetrically distributed directions in the plane at low
temperatures (compare Fig.\ \ref{figFlucJneg}). The latter prediction
is indeed verified by the histogram of the angles of the in-plane component
of the spins $\phi_i$ at low temperatures shown in Fig.\ \ref{figStatPhi}.
Note that the histogram is rather flat for the smaller lattices (in particular
the $N = 6\times 6$ lattice) and sharpens noticeably as the lattice
size increase to $N = 36 \times 36$ (the largest lattice which we have
considered in this context). The fact that the lock-in occurs only on large
lattices can be attributed to the replacement of the sum over $\vec{k}$ in
(\ref{FlucJneg}) by an integral being a good approximation only for large
lattices.

\begin{figure}[tb!]
\begin{center}
\includegraphics[width=\columnwidth]{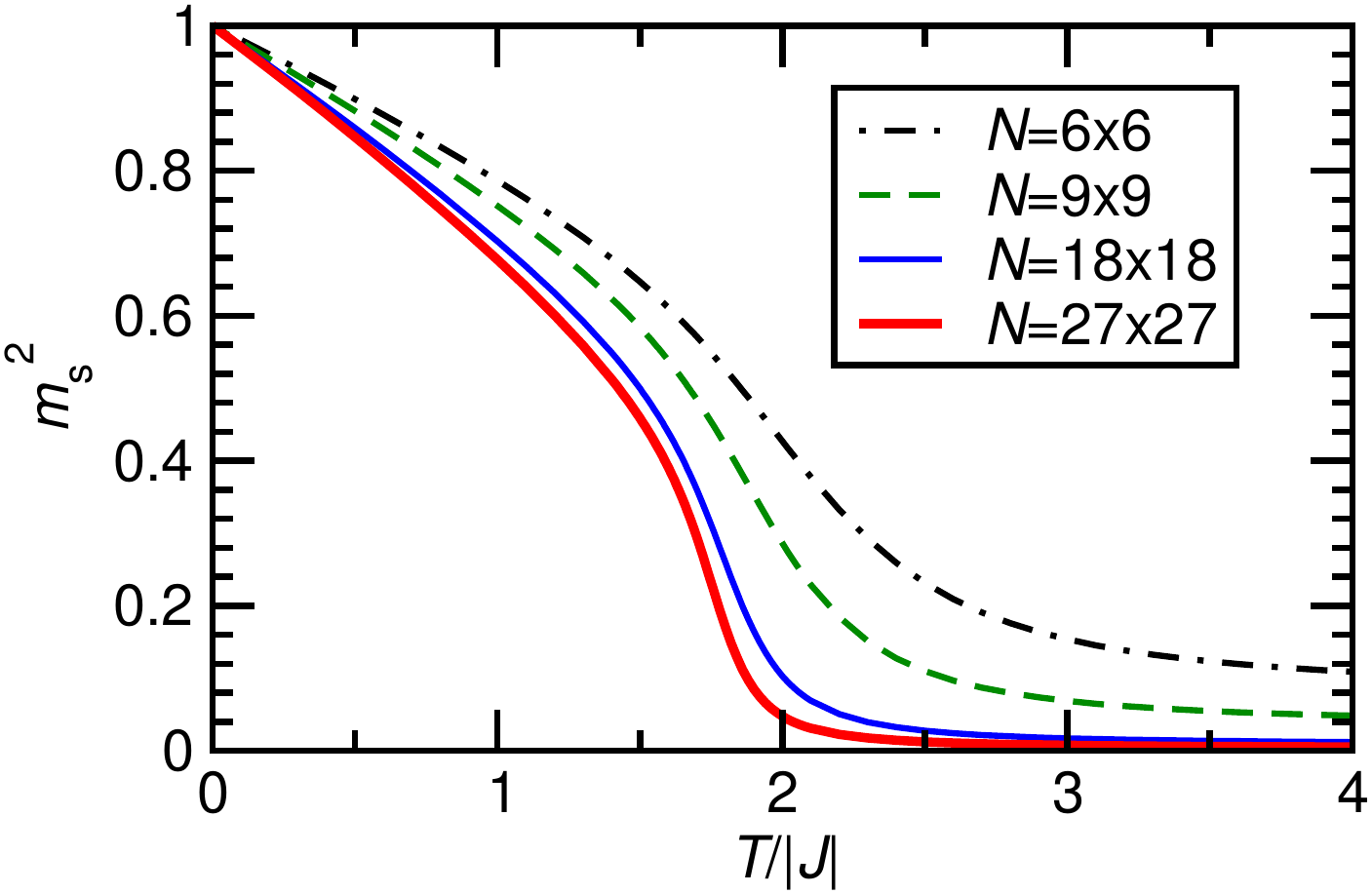}
\end{center}
\caption{(Color online) Square of the sublattice order parameter
$m_{\rm s}^2 =
\left\langle \vec{M}_{\rm s}^2 \right\rangle$ for the classical
model with $J < 0$.
Error bars do not exceed the width of the lines.}
\label{figMsublNeg}
\end{figure}

To test for the expected three-sublattice order, we introduce
the sublattice order parameter
\begin{equation}
\vec{M}_{\rm s} = {3 \over N} \sum_{i \in {\cal L}} \vec{S}_i \, ,
\label{defMsubl}
\end{equation}
where the sum runs over one of the three sublattices ${\cal L}$
of the triangular lattice. Fig.\ \ref{figMsublNeg} shows the behavior
of the square of this sublattice order parameter for $J<0$.
One observes that the sublattice order parameter indeed increases
for $T < 2\,\abs{J}$ and goes indeed to $m_{\rm s}^2 = 1$ for $T \to 0$,
as expected
for a three-sublattice ordered state. Inclusion of larger lattices
(up to $N=90\times90$) allows one to restrict the ordered phase to
$T \lesssim 1.7\,\abs{J}$. However, more accurate estimates for the
transition temperature can be obtained in a different manner.

A useful quantity to determine the transition into an ordered
state accurately
is the `Binder' cumulant\cite{Binder81a,Binder81b,LB00} associated to
the order parameter (\ref{defMsubl}) via
\begin{equation}
U_{\rm s} = 1 - {3 \, \left\langle \vec{M}_{\rm s}^4 \right\rangle \over
5 \, \left\langle \vec{M}_{\rm s}^2 \right\rangle^2} \, .
\label{defBinder}
\end{equation}
We have chosen the prefactor in (\ref{defBinder}) such that
$U_{\rm s}=0$ for a Gaussian distribution around zero of the order
parameter $P(\vec{M}_{\rm s}) = \left({c \over \pi}\right)^{3/2} \,
\exp\left(-c\,\vec{M}_{\rm s}^2\right)$. Such a distribution is expected
at high temperatures, and we expect $U_{\rm s} \to 0$ for
$T \gg \abs{J}$. Conversely, in a perfectly ordered state
one will have $\left\langle \vec{M}_{\rm s}^4 \right\rangle =
\left\langle \vec{M}_{\rm s}^2 \right\rangle^2$ such that $U_{\rm s} = 2/5$
in this case. Hence, for an ordered state we expect $U_{\rm s} \approx 0.4$
for $T < T_c$.

\begin{figure}[tb!]
\begin{center}
\includegraphics[width=\columnwidth]{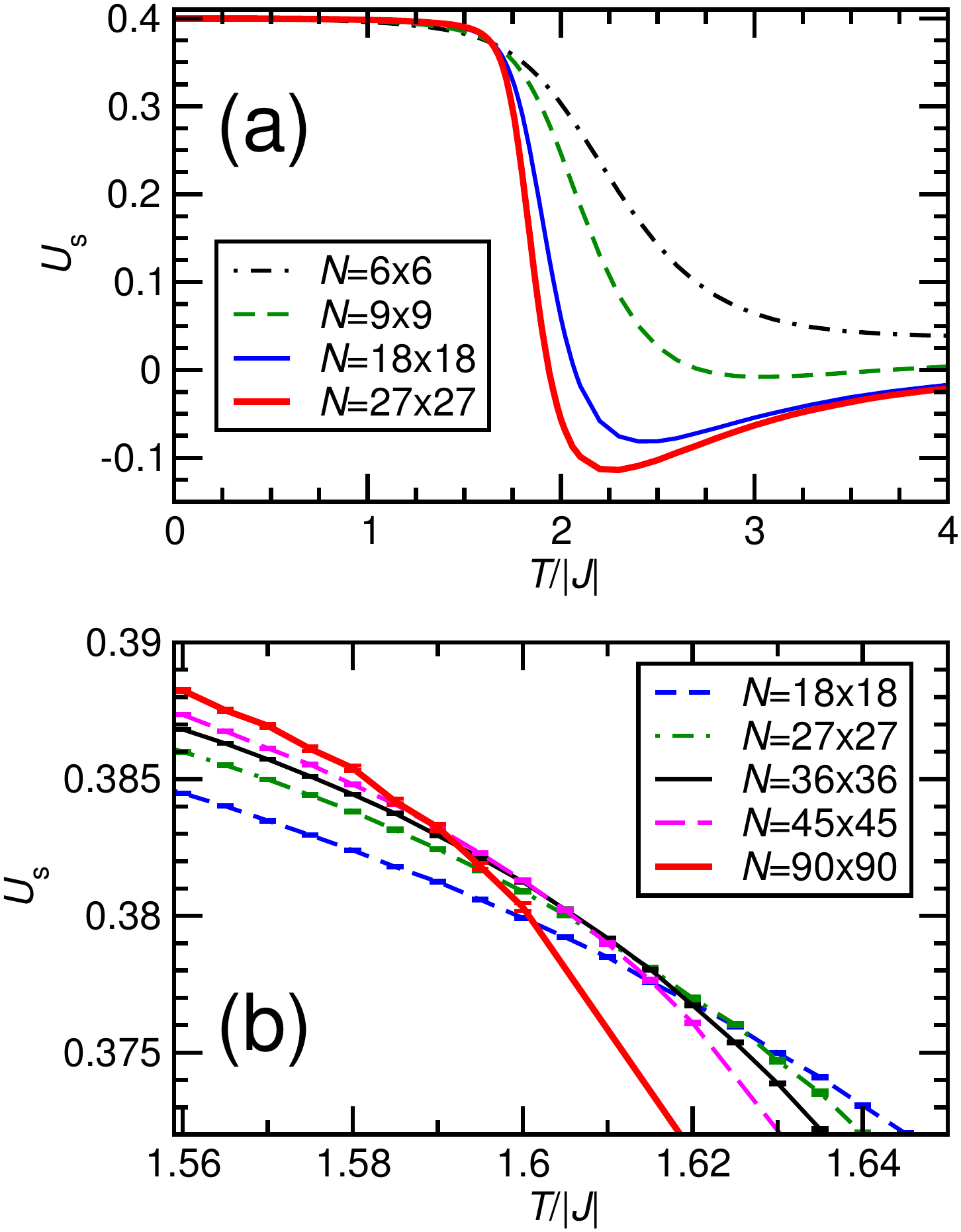}
\end{center}
\caption{(Color online) Binder cumulant $U_{\rm s}$ for the classical
model with $J < 0$:
(a) global behavior,
(b) in the vicinity of the critical temperature and for bigger lattices.
Error bars in panel (a) do not exceed the width of the lines.}
\label{figBinderNeg}
\end{figure}

Fig.~\ref{figBinderNeg} shows
results from classical Monte-Carlo simulations for the Binder cumulant
$U_{\rm s}$ of the system with $J<0$. First, the broad temperature
range shown in Fig.~\ref{figBinderNeg}(a) confirms that indeed
$U_{\rm s} \approx 0.4$ in the ordered low-temperature phase and
$U_{\rm s} \approx 0$ for high temperatures.
The transition temperature can now be accurately extracted from the crossings of
the Binder cumulants at different sizes $N$.\cite{Binder81a,Binder81b,LB00}
For this purpose, Fig.~\ref{figBinderNeg}(b) zooms in to the relevant
temperature range, including bigger system sizes $N$. Although the crossings
between any pair of system sizes $N_1$ and $N_2$ fall into a narrow temperature
window, there still remains a small residual dependence on the sizes $N_1$
and $N_2$ considered. In order to perform an extrapolation $N\to\infty$,
we have analyzed the crossings between neighboring system sizes
$N_2 > N_1 \ge 9\times9$. This leads to the estimate
\begin{equation}
\frac{T_c^\Neg}{\abs{J}} =  1.566 \pm 0.005 
\label{TcNeg}
\end{equation}
for the thermodynamic limit $N\to\infty$.

\subsection{Nature of the phase transition for $J<0$}

Having determined the transition temperature for $J<0$, one would like
to clarify the universality class of the phase transition.

\begin{figure}[tb!]
\begin{center}
\includegraphics[width=\columnwidth]{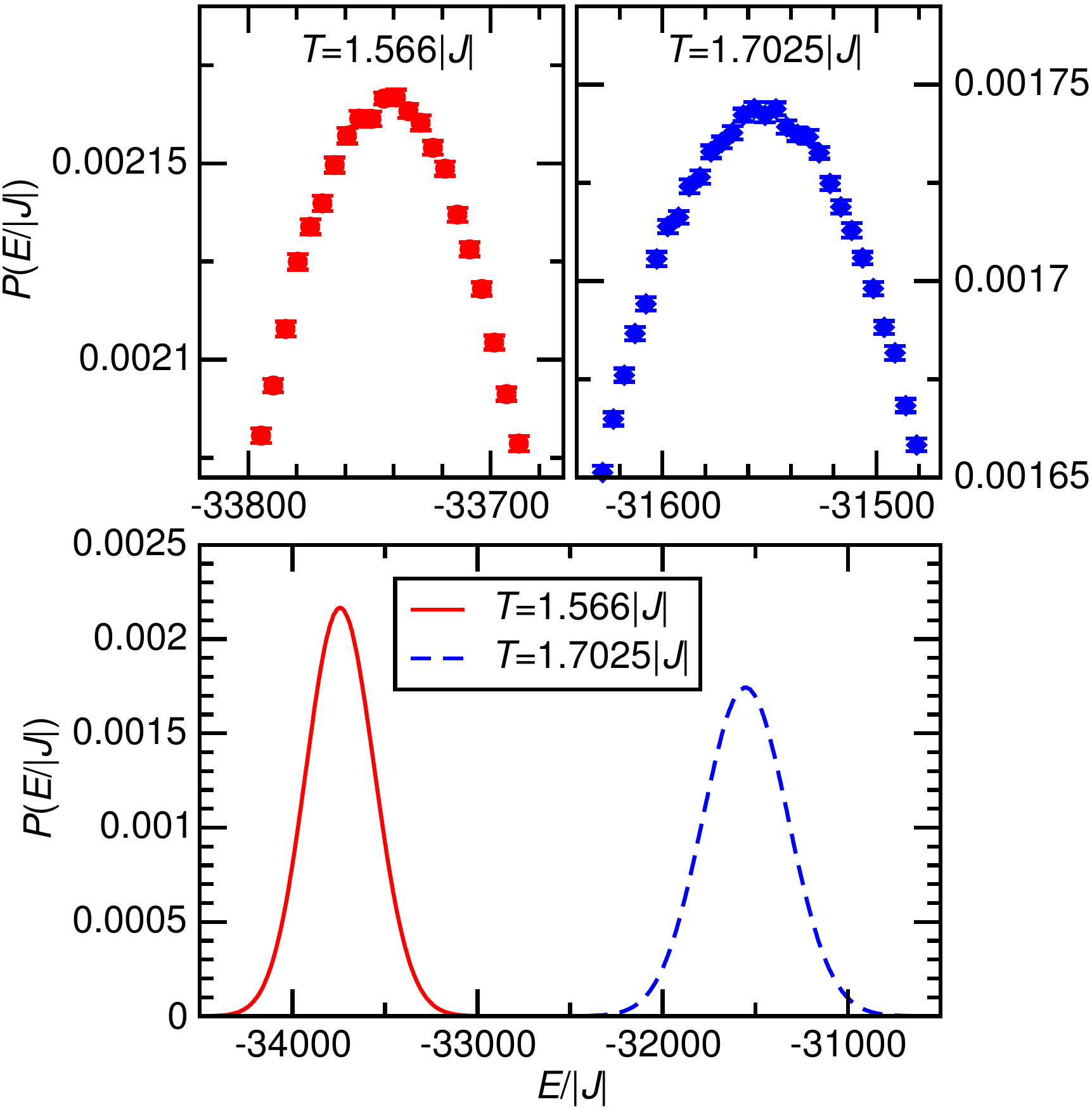}
\end{center}
\caption{(Color online) Probability to find a state with
energy $E/\abs{J}$ on the $N=90\times90$ lattice for $J < 0$
at two selected temperatures: $T/\abs{J} = 1.566$ (left)
and $1.7025$ (right).
Error bars in the lower panel do not exceed the width of the lines.}
\label{figHistNeg}
\end{figure}

On the one hand, there is no evidence for any latent heat in the
specific heat at $T_c^\Neg$, see Fig.~\ref{figCzoomed}(a), {\it i.e.}, the
ordering transition appears to be continuous for $J<0$.
On the other hand, a negative dip in the Binder cumulant for
$T>T_c$, as observed in Fig.~\ref{figBinderNeg}(a) is sometimes
taken as evidence for a first-order transition
(see, e.g., Ref.~\onlinecite{OK10}). In order to distinguish better
between the two scenarios we use histograms of the energy $E$ of the
microstates realized in the Monte-Carlo
procedure.\cite{P:challa86,P:berg91,P:borgs92}
We have collected such histograms for several system sizes and temperatures.
Fig.~\ref{figHistNeg} shows two representative cases on the $N=90\times90$
lattice, namely $T=1.7025\,\abs{J}$ which corresponds to the maximum
of the specific heat for the $90\times90$ lattice (compare
Fig.~\ref{figCzoomed}(a)) and  $T=1.566\,\abs{J}$, the estimated critical
temperature of the infinite system, see
Eq.~(\ref{TcNeg}). We always find bell-shaped almost Gaussian distributions,
which are characteristic for a continuous transition. We never observed
any signatures of a splitting of this single peak into two, as would
be expected for a first-order transition.\cite{P:challa86,P:berg91,P:borgs92}
Hence, the transition appears to be continuous and we will now try to
characterize its universality class further in terms of critical exponents.

\begin{figure}[tb!]
\begin{center}
\includegraphics[width=\columnwidth]{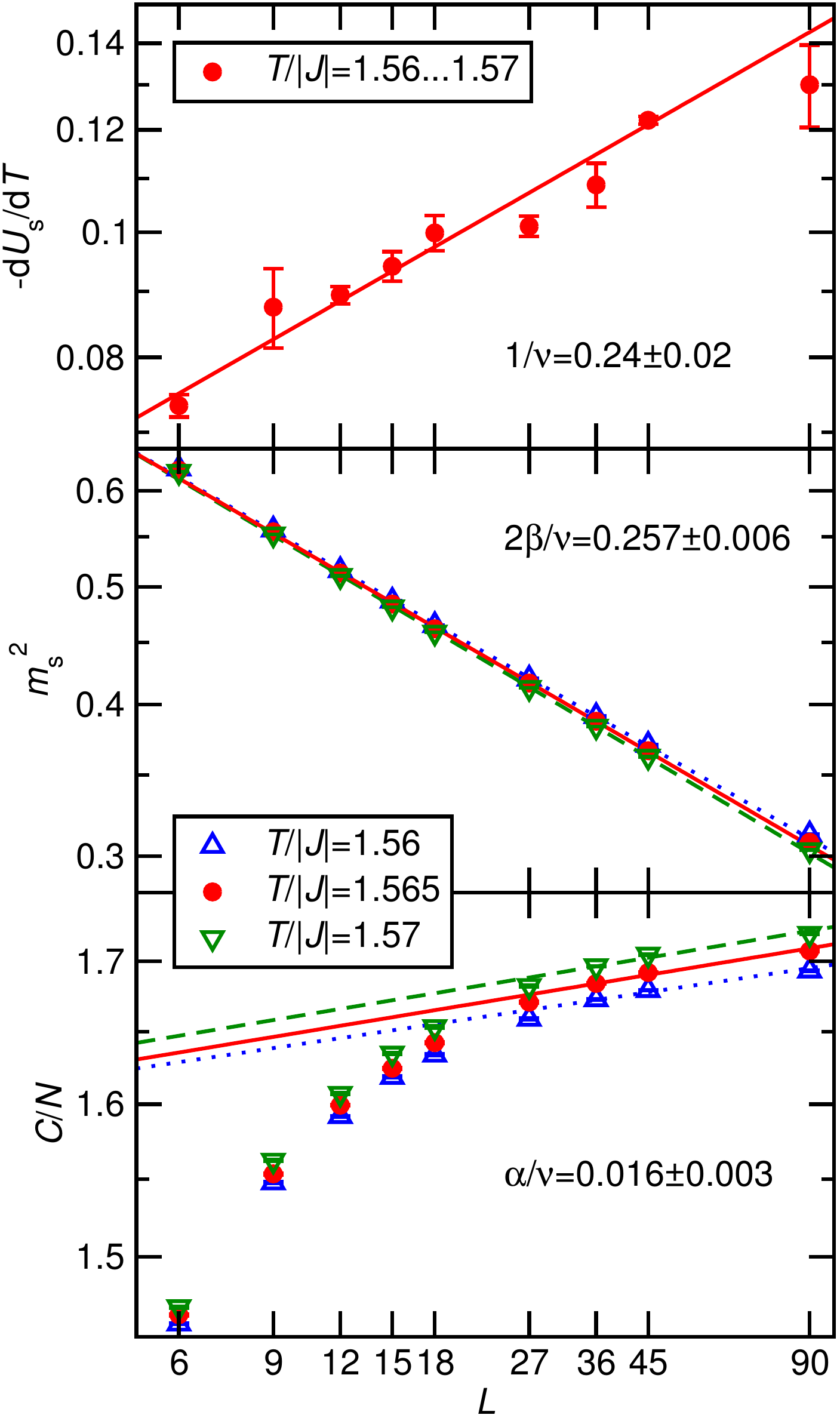}
\end{center}
\caption{(Color online) Scaling of different quantities with linear
size $L$ for $L\times L$ lattices, $J<0$, and close to the critical
temperature $T_c^{\Neg}$. The slope of the Binder cumulant yields
the correlation length exponent $\nu$ (top panel), the sublattice
magnetization $m_{\rm s}$ yields the exponent $\beta$, and the
specific heat $C$ yields the exponent $\alpha$.
Lines show the  fits which have been used to estimate the exponents.
Note that the scale is double-logarithmic in all three panels.}
\label{figExponents}
\end{figure}

We start with the correlation length exponent $\nu$ which can be extracted
from the finite-size behavior of the Binder cumulant:
close to $T_c$, the Binder cumulant should scale
with the {\em linear} size of the system $L$ as\cite{Binder81a,Binder81b,LB00}
\begin{equation}
{{\rm d} \, U_{\rm s} \over {\rm d} T} \approx
a \, L^{1/\nu} \, \left(1 + b \, L^{-w}\right) \, .
\label{defInu}
\end{equation}
Fig.~\ref{figBinderNeg}(b) shows that there is very little curvature
in the Binder cumulants $U_{\rm s}$ as a function of temperature $T$
close to the estimated critical temperature (\ref{TcNeg}). Therefore
${{\rm d} \, U_{\rm s}}/{{\rm d} T}$ can be extracted without much
sensitivity to the error of (\ref{TcNeg}). The result is shown by the
symbols in the top panel of Fig.~\ref{figExponents}. Now we can determine
$\nu$ by fitting this to (\ref{defInu}). Since inclusion of the correction
term renders the fit unstable,
we use only the leading term ({\it i.e.}, we set $b=0$ in (\ref{defInu})).
A fit (which is shown by the line in the top panel of Fig.~\ref{figExponents})
then yields
\begin{equation}
\frac{1}{\nu} = 0.24 \pm 0.02 \, .
\label{resInuJneg}
\end{equation}

We now turn to the order parameter exponent $\beta$ which can be
extracted from the finite-size behavior of the
sublattice order parameter $m_{\rm s}$.
The sublattice order parameter should have a scaling behavior
(see for example Refs.\ \onlinecite{Landau76,LB00})
\begin{equation}
m_{\rm s}^2 =
\left\langle \vec{M}_{\rm s}^2 \right\rangle =
L^{-2 \, \beta/\nu} \, {\cal M}_2\left(\left(1 - {T \over T_c}\right) \,
L^{1/\nu}\right) \, .
\label{Mscale}
\end{equation}
Specialization of (\ref{Mscale}) to $T=T_c$ yields
\begin{equation}
\left.\left\langle \vec{M}_{\rm s}^2 \right\rangle\right\vert_{T=T_c}
= L^{-2 \, \beta/\nu} \, {\cal M}_2\left(0\right) \, .
\label{MscaleTc}
\end{equation}
The middle panel of Fig.~\ref{figExponents} shows the Monte-Carlo
results for $m_{\rm s}^2$ at three temperatures which cover the
estimate (\ref{TcNeg}) for $T_c^\Neg$ and its error bars.
The fits of these results to (\ref{MscaleTc}) which are shown by the
lines in the middle panel of Fig.~\ref{figExponents} lead to
\begin{equation}
\frac{2 \, \beta}{\nu} = 0.257 \pm 0.006 \, .
\label{res2betaInuJneg}
\end{equation}

Finally, we turn to the specific heat exponent $\alpha$. We proceed in
the same manner as for the order parameter exponent $\beta$ and make
again a scaling ansatz for the specific heat:\cite{Landau76,LB00}
\begin{equation}
\left. C \right\vert_{T=T_c}
= L^{\alpha/\nu} \, {\cal C}\left(0\right) \, .
\label{CscaleTc}
\end{equation}
The lower panel of Fig.~\ref{figExponents} shows the specific heat results
at the estimate (\ref{TcNeg}) for $T_c^\Neg$. One observes that this does
not follow a power law very well. Indeed, it is known that
non-scaling contributions to the specific heat can
be important.\cite{Landau76} However, including a constant
in the ansatz (\ref{CscaleTc}) does not lead to a stable fit. We therefore
fit only the data  for $L=27$, $36$, $45$, and $90$ (lines in the
lower panel of Fig.~\ref{figExponents}). This procedure leads to the estimate
\begin{equation}
\frac{\alpha}{\nu} = 0.016 \pm 0.003 \, .
\label{resAlphaInuJneg}
\end{equation}
Note that the error bar includes just the error of the fit. In view
of the deviations from a simple power law, this is probably too optimistic.
Indeed, $\alpha$ could very well be (slightly) negative, as is suggested
by the fact that the maximal value of $C$ in Fig.\ \ref{figCzoomed}(a)
decreases when the system size increases from $N=45\times45$ to $90\times90$.

Even if the error bars in (\ref{TcNeg}), (\ref{resInuJneg}), and
(\ref{resAlphaInuJneg}) should be too optimistic, it
remains safe to conclude that we find a rather large correlation length
exponent $\nu \gtrsim 3$.
It should be noted that in combination with
a specific heat exponent $\alpha \approx 0$, we then find that 
the hyperscaling relation\cite{LB00}
\begin{equation}
d\,\nu = 2 - \alpha
\label{hyperscaling}
\end{equation}
(with the spatial dimension $d=2$) is strongly violated.
On the other hand, we could use the relation (\ref{hyperscaling})
to estimate $\alpha$, in particular if we expect it to be negative
(compare, e.g., Ref.~\onlinecite{PPR99} for a similar situation).
Insertion of (\ref{resInuJneg}) into (\ref{hyperscaling}) yields
a very negative exponent $\alpha/\nu =-1.52 \pm 0.04$. Again, this
reflects the large exponent $\nu$. In fact, a large correlation length
exponent $\nu \approx 4$ has been found in other two-dimensional disordered
systems.\cite{MC02,PHP06} However, in those cases the large exponent
corresponds to approaching the critical point via a fine-tuned direction in
a two-dimensional parameter space and there is a second, substantially smaller
correlation length exponent.\cite{MC02,PHP06} Thus, we are left
with not completely unreasonable, but definitely highly unusual values
of the critical exponents $\nu$ and $\alpha$.

\subsection{Critical temperature for $J>0$}

\begin{figure}[tb!]
\begin{center}
\includegraphics[width=\columnwidth]{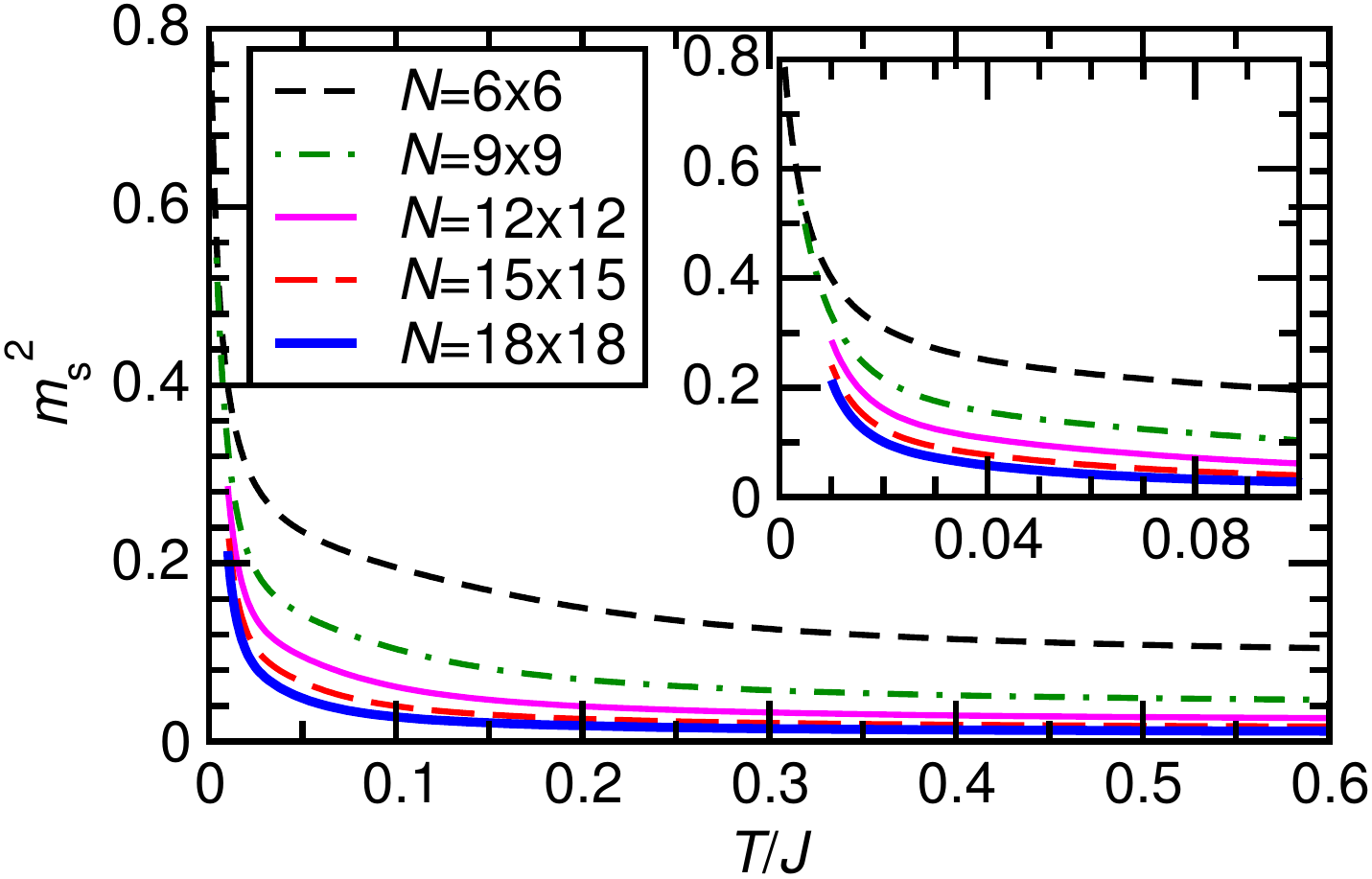}
\end{center}
\caption{(Color online) Average squared sublattice magnetization for the
classical model with $J>0$. Error bars are of the order of the lines'
width in this graph. Inset: average squared sublattice magnetization
in the low-temperature region.}
\label{figsublatpos}
\end{figure}

As mentioned earlier, for $J>0$, the specific heat alone is only mildly
conclusive regarding the existence of a low-temperature phase transition
to a three-sublattice ordered state. This statement requires to be
supported by the analysis of other observables. The sublattice
magnetization (\ref{defMsubl}) will
tell us whether significant order is developing
in the low-temperature region or not. Our results shown in Fig.\
\ref{figsublatpos} show that three-sublattice order is indeed developing,
although an appreciable order develops only at temperatures that are so
low that they become increasingly difficult to access with increasing
system size. To take a closer look at the low-temperature ordered state,
we took some snapshots of the system during the simulation for
$N=12\times 12$ spins. A typical configuration is reproduced in Fig.\
\ref{figSnapshotPos}. While the global structure corresponds indeed to a
$120^{\circ}$ three-sublattice ordered state, we also observe the presence
of defects. The presence of these defects is neither a surprise nor in
contradiction with the existence of the phase transition, as they are
in fact necessary ingredients of the order-by-disorder mechanism.
Note also that the spins in Fig.\ \ref{figSnapshotPos} lie
essentially in the $x-y$-plane and --up to small fluctuations--
are aligned with the lattice.

\begin{figure}[tb!]
\begin{center}
\includegraphics[width=\columnwidth]{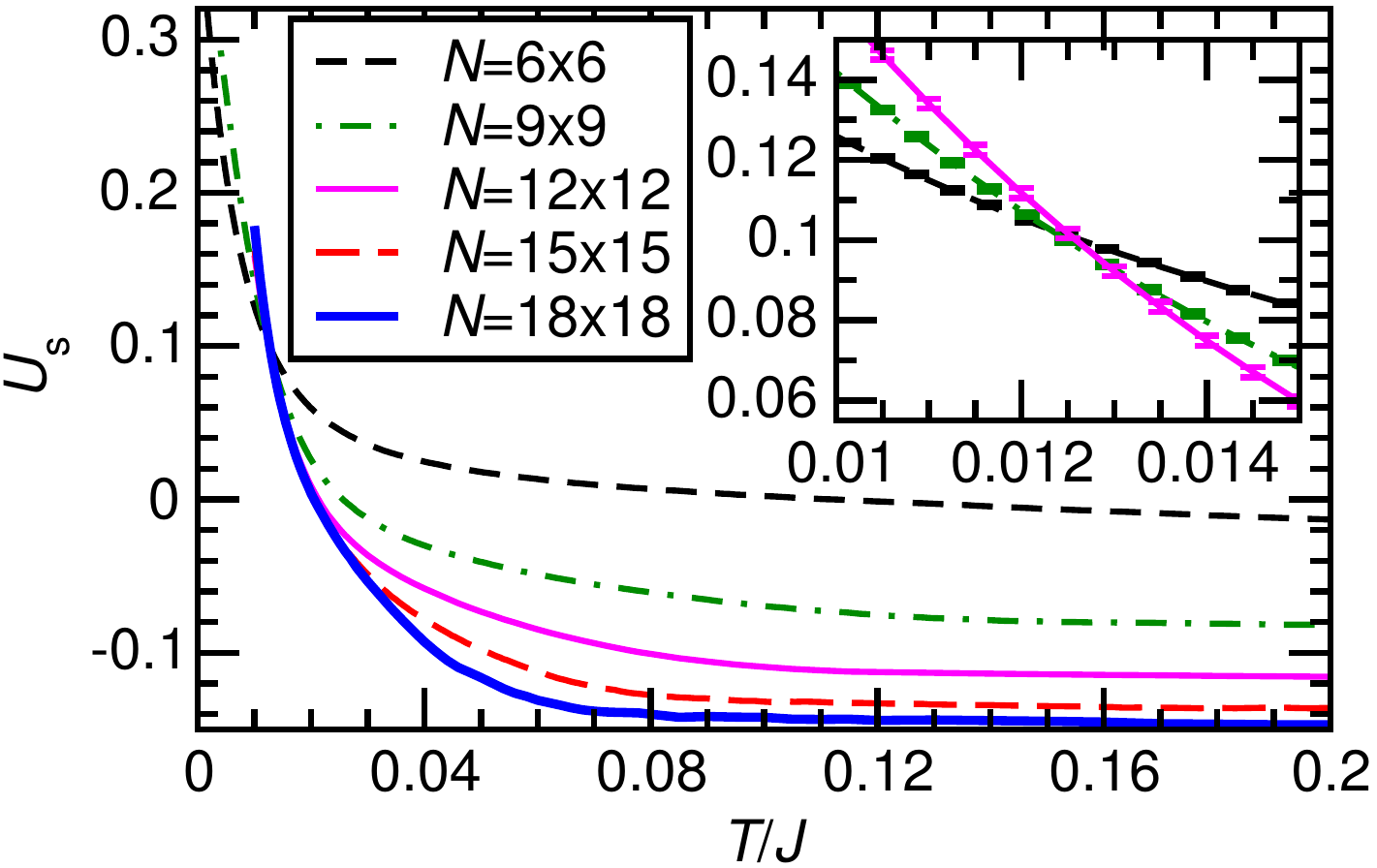}
\end{center}
\caption{(Color online) Binder cumulant
for the classical model with $J>0$  in the low-temperature region. Error bars are of the order of the lines' width in this graph. Inset: Binder cumulant
for $6\times 6$, $9\times 9$, and $12\times 12$ spins zoomed around the crossing region.
}
\label{figbinderpos}
\end{figure}

As for $J<0$, the Binder cumulant (\ref{defBinder})
allows us both to further support our conclusions concerning
the low-temperature ordered state and to obtain an estimate of $T_c^\Pos$.
Fig.\ \ref{figbinderpos} shows that all the curves for the different system sizes cross in a region around $T\approx 0.015 J$.
First, this is a strong argument in favor of the existence of the
ordering transition.
We used the smallest three system sizes ($N=6\times 6$, $9\times 9$, and $12\times 12$) for which we have the best statistics to
obtain an estimate for the transition temperature:
\begin{equation} 
	\frac{T_c^\Pos}{J} = 0.0125\pm 0.0009.
\end{equation} 
The error bars on the data are unfortunately too large to get precise values for the critical exponents and thus prevent us
from investigating the nature and the universality class of the transition.
However, the fact that the low-temperature ordered state breaks the
same symmetries irrespectively of the sign of $J$ suggests that the universality
class for $J>0$ is the same as for $J<0$.

\section{Discussion and Conclusions}

\label{sec:Concl}

Although the Hamiltonian (\ref{eqH}) may seem unusual in the context of
frustrated magnetism, it is instructive in many respects and illustrates the
rich phenomenology often present in this subject. Either seen as the
strong trimerization limit of the \kagome\ lattice of spin $1/2$ in a
magnetic field, or as a possible illustration of an orbital model,
\cite{NuFr05,DBM05,TI07,WJ08,SWT09,WJ09,TOH10,WJL10,OH11}
the underlying physics associated to this Hamiltonian is
extremely interesting for both signs of the coupling constant $J$.

In the quantum case (for spin $1/2$) we show, by studying the low-energy
spectra using the Lanczos method, that a thermodynamic gap of the order of $3 \, \vert
J\vert$ is present for $J < 0$, while for $J > 0$ the gap, if present,
would be at most of the order of $0.02 \,J$. The six-fold degeneracy
of the would-be ordered ground state which is predicted by semiclassical
considerations is not observed in our numerical results, probably due
to the small lattices considered. These results illustrate very well how
the deep quantum ($S = 1/2$) regime differs from the large $S$ spin-wave
predictions. The specific heat curves point to a phase transition around
$T \approx \vert J \vert $ for $J < 0$, while a lower temperature peak
shows up in the positive $J$ case. This last peak could be due to an
ordering phase transition at a very low temperature $T_c \le J/100$. In
both cases one is tempted to envisage a finite-temperature phase
transition whose nature could be understood by the analysis of the
classical model.

The analysis of the classical model has turned out to be also quite
interesting and instructive. For $J<0$ the lowest-energy configuration
consists in an in-plane antiferromagnetic arrangement of the spins with
given chirality accompanied by a `spurious' continuous rotational degeneracy
which does not correspond to any symmetry of the Hamiltonian. This pseudo
degeneracy is lifted by entropy at finite temperature giving rise to an
ordering at low temperature as observed by Monte Carlo data which locate
the transition temperature at $T_c /\vert J \vert = 1.566 \pm 0.005$.
Inspection of the histograms of the energy close to the transition temperature
gave no evidence of a first-order transition. Hence, we analyzed it within
the scenario of a continuous transition and estimated  
unusual values for the critical exponents $\alpha$ and $\nu$,
strongly violating the hyperscaling relation. It should nevertheless be mentioned
that we cannot exclude the existence of a crossover scale which exceeds
the lattice sizes accessible to us. The fact that lock-in of the spin
components to the lattice requires a certain length scale may point in
this direction. An unambiguous determination of the universality class of
the transition would require improved methods. A first possibility is to
restrict the degrees of freedom to the in-plane configurations\cite{hfm2006}
which are realized in the low-temperature limit. Even more efficiency
could be gained by additionally restricting each spin variable to the 6
spin directions which are stabilized in the zero-temperature limit.
However, it remains to be investigated whether the second modification
changes the universality class of the transition.

For $J>0$ the situation is even more interesting. Although a `spurious'
rotational degeneracy is also present for the antiferromagnetic $120^\circ$
configuration (with the opposite chirality than the one for $J<0$), the
manifold of
lowest-energy configurations is more complex.  There exist
local discrete `flips' of triangles which bring one from the homogeneous
antiferromagnetic lowest-energy configuration to another configuration with
the same energy. The mechanism that gives rise to ordering is again
understood by analyzing the entropic spectra over each of these
configurations. The homogeneous antiferromagnetic configuration has a
whole branch of soft modes in its classical spin-wave spectrum. Flipping
one triangle to jump to another lowest-energy configuration also
destroys one soft mode. One is then left with a scenario that can
be understood with an Ising-type low-temperature expansion
picture of the system, where each `flippable' triangle plays the r\^ole of
an Ising spin. The difference is in the fact that flipping one spin on an
otherwise perfectly ordered background costs no energy but an entropy, or if
one wishes a temperature-dependent pseudo energy. An ordering transition
will also take place, as in a normal energetic system, but at a much
smaller temperature. This transition temperature is observed in the
Monte-Carlo analysis to be at around $T_c /\vert J \vert \approx 0.0125$,
two orders of magnitude smaller than in the $J<0$ case.

\begin{figure}[tb!]
\includegraphics[width=\columnwidth]{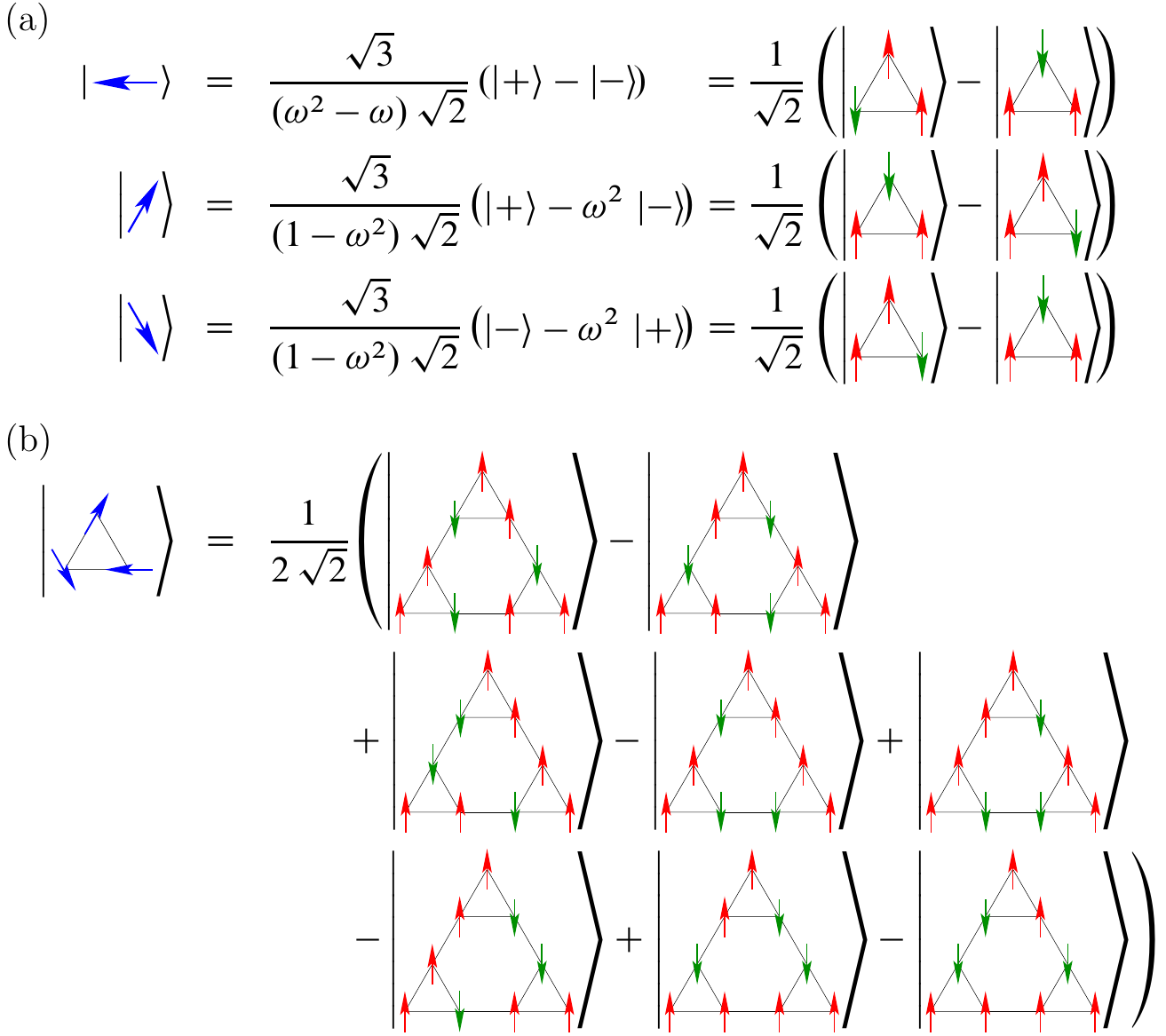}
\caption{(Color online)
(a) Identification of chirality pseudo-spin states of a triangle with
spin configurations on a triangle for the underlying \kagome\ lattice.
(b) Expression of the $120^{\circ}$ ordered state on a triangle in the
effective model in terms of spin configurations of the corresponding 9 sites
in the underlying \kagome\ lattice. Note that chirality spins for the
effective model lie in the $x$-$y$ chirality plane whereas spins on the
\kagome\ lattice point along the $z$-axis.
}
\label{figM1o3state}
\end{figure}

For $J>0$, one may also wonder how the $120^\circ$ ordered state of the 
model (\ref{eqH}) relates to the structure of the magnetization $1/3$ 
state of the homogeneous \kagome\ lattice.\cite{kagomeus,kagomeus1} In 
order to address this question, we need to associate a variational wave 
function to the classical $120^\circ$ ordered state which is indicated in 
Fig.~\ref{figSnapshotPos}. First, we associate quantum wave functions to 
the three classical spin directions as in Fig.~\ref{figM1o3state}(a). The 
phase factors are chosen in order to yield a convenient representation in 
terms of spin configurations of a triangle after insertion of 
Fig.~\ref{figLattice}(b) for the chirality pseudo spins. Insertion into 
the $120^\circ$ wavefunction for a triangle of the triangular lattice on 
the left side of Fig.~\ref{figM1o3state}(b) then yields the expression in 
terms of the 8 spin configurations of a nine-site unit of the underlying 
\kagome\ lattice shown on the right side of Fig.~\ref{figM1o3state}(b). 
Note that the two terms on the first line of the right side of 
Fig.~\ref{figM1o3state}(b) amount exactly to the variational wave function 
for the magnetization $1/3$ state of the homogeneous \kagome\ 
lattice,\cite{kagomeus,kagomeus1} as it follows from a mapping to a 
quantum-dimer model on the honeycomb lattice.\cite{MS01} Thus, the present 
results for the strongly trimerized \kagome\ lattice may be smoothly 
connected to the 1/3 plateau state of the homogeneous \kagome\ lattice.

To conclude, the model (\ref{eqH}) has turned out to be a very interesting 
laboratory to understand the emergence of a hierarchy of energy scales 
originating from different levels of order by disorder. The emergence of 
such a hierarchy in a classical model is related to similar hierarchies in 
quantum systems as for example the huge difference between the magnetic 
and non-magnetic gaps in the \kagome\ spin $1/2$ system at magnetization 
$1/3$.\cite{kagomeus} Moreover, if the transitions observed in this work 
can be confirmed to be continuous, the exponents will probably correspond 
to exotic models, like parafermionic conformal field theories.\cite{FZ}

\acknowledgments

We are grateful to P.~C.~W.\ Holdsworth, H.~G.\ Katzgraber, F.\ Mila, and 
M.~E.\ Zhitomirksy for useful discussions, comments and suggestions.  
This work has been supported in part by the European Science Foundation 
through the Highly Frustrated Magnetism network. D.C.C.\ is partially 
supported by CONICET (PIP 1691) and ANPCyT (PICT 1426). Furthermore, A.H.\ 
is supported by the Deutsche Forschungsgemeinschaft through a Heisenberg 
fellowship (Project HO~2325/4-2).

\appendix

\section{Relation to the Heisenberg model on the trimerized \kagome\
lattice}

\label{sec:AppDeriv}

The Hamiltonian (\ref{eqH}) has already been derived several times
in the literature.\cite{sub95,SBCEFL04,Z05} Nevertheless, for completeness
we also give a derivation.

We start from the interaction between the triangles of the spin-1/2
Heisenberg model on a trimerized \kagome\ lattice
\begin{eqnarray}
H_{\rm int} &=& J_{\rm int} \sum_{\langle i, j\rangle}
  \vec{\cal S}_{{\rm A},i} \cdot \vec{\cal S}_{{\rm B},j} \label{eq:HHeisInt} \\
  &=& J_{\rm int} \sum_{\langle i,j\rangle}
  \left\{\frac{1}{2} \left(
  {\cal S}_{{\rm A},i}^+ \, {\cal S}_{{\rm B},j}^-
  + {\cal S}_{{\rm A},i}^- \, {\cal S}_{{\rm B},j}^+\right)
  + {\cal S}_{{\rm A},i}^z \, {\cal S}_{{\rm B},j}^z \right\} \, , \nonumber
\end{eqnarray}
where the sum over $\langle i, j\rangle$ runs over the nearest-neighbor
pairs of triangles $i$ and $j$ in Fig.~\ref{figLattice}(a) and the corners
of the triangle A and B have to be chosen such as to match the connecting
bond. The $\vec{\cal S}_{{\rm A},i}$ are physical spin-1/2 operators.

In first order and for $N$ triangles, we need to compute the matrix elements
of (\ref{eq:HHeisInt}) between all $2^N$ combinations of the states in
Fig.~\ref{figLattice}(b). Note that the expectation values of the
operators acting on different triangles factorize. Since in the present case
magnetization is fixed in each triangle to 1/3, matrix elements
of ${\cal S}_{{\rm A},i}^\pm$ vanish, thus simplifying the derivation
considerably.

For the lower left corner of a triangle $i$ we find
\begin{eqnarray}
\pmatrix{
\langle+|{\cal S}_{{\rm L},i}^z|+\rangle & \langle+|{\cal S}_{{\rm L},i}^z|-\rangle \cr
\langle-|{\cal S}_{{\rm L},i}^z|+\rangle & \langle-|{\cal S}_{{\rm L},i}^z|-\rangle \cr
}
&=&
\pmatrix{
1/6 & -\omega^2/3 \cr
-\omega/3 & 1/6 \cr
}_i \nonumber\\
&=& \frac{1}{3}\left(\frac{1}{2} -T_i^C\right) \, ,
\label{eq:RepSzL}
\end{eqnarray}
for the lower right corner
\begin{eqnarray}
\pmatrix{
\langle+|{\cal S}_{{\rm R},i}^z|+\rangle & \langle+|{\cal S}_{{\rm R},i}^z|-\rangle \cr
\langle-|{\cal S}_{{\rm R},i}^z|+\rangle & \langle-|{\cal S}_{{\rm R},i}^z|-\rangle \cr
}
&=&
\pmatrix{
1/6 & -1/3 \cr
-1/3 & 1/6 \cr
}_i \nonumber\\
&=& \frac{1}{3}\left(\frac{1}{2} -T_i^A\right) \, ,
\label{eq:RepSzR}
\end{eqnarray}
and finally for the top corner
\begin{eqnarray}
\pmatrix{
\langle+|{\cal S}_{{\rm T},i}^z|+\rangle & \langle+|{\cal S}_{{\rm T},i}^z|-\rangle \cr
\langle-|{\cal S}_{{\rm T},i}^z|+\rangle & \langle-|{\cal S}_{{\rm T},i}^z|-\rangle \cr
}
&=&
\pmatrix{
1/6 & -\omega/3 \cr
-\omega^2/3 & 1/6 \cr
}_i \nonumber\\
&=& \frac{1}{3}\left(\frac{1}{2} -T_i^B\right) \, .
\label{eq:RepSzT}
\end{eqnarray}
Using (\ref{eq:RepSzL})--(\ref{eq:RepSzT}) for the matrix
elements of (\ref{eq:HHeisInt}), we find the effective Hamiltonian
\begin{eqnarray}
H_{\rm eff.} &=& \frac{J_{\rm int}}{9} \, \left\{\sum_{\langle i,j\rangle}
  \left(\frac{1}{2} - T^A_i \right) \left(\frac{1}{2} - T^C_j \right)
  \right.\nonumber \\
&& \qquad + \sum_{\langle\langle k,j \rangle\rangle}
   \left(\frac{1}{2} - T^A_k \right) \left(\frac{1}{2} -T^B_j  \right)
   \nonumber \\
&& \qquad + \left.\sum_{[[ k,i ]]}
  \left(\frac{1}{2} - T^C_k \right) \left(\frac{1}{2} - T^B_i \right)
  \right\} \, ,
\label{eqHeff}
\end{eqnarray}
where the three sums run over the different bond directions as sketched
in Fig.~\ref{figLattice}(a). Since the sum over roots of unity vanishes,
we have $T^A_i + T^B_i + T^C_i = 0$. Hence, Eq.~(\ref{eqHeff}) can be
rewritten as
\begin{eqnarray}
H_{\rm eff.} &=& \frac{J_{\rm int}}{9} \, \left\{\sum_{\langle i,j\rangle} T^A_i T^C_j
+ \sum_{\langle\langle k,j \rangle\rangle} T^A_k T^B_j
+ \sum_{[[ k,i ]]} T^C_k T^B_i \right\}
\nonumber \\
&& + \frac{N\,J_{\rm int}}{12} \, .
\label{eqHeffR}
\end{eqnarray}
Up to an additive constant, this is nothing but Eq.~(\ref{eqH}) with
$J=J_{\rm int}/9$. The intra-triangle coupling needs to be chosen positive
in order for the two states shown in Fig.~\ref{figLattice}(b)
to be ground states of a triangle. Accordingly, it is natural to also
choose the inter-triangle coupling $J_{\rm int}$ positive, {\it i.e.}, $J > 0$.

Very similar arguments can be applied, e.g., to spinless fermions with
nearest-neighbor repulsion,\cite{SBCEFL04} leading to the same
effective Hamiltonian.

\end{document}